\documentclass[aip,reprint]{revtex4-1}

\providecommand{\tabularnewline}{\\}
\usepackage{epstopdf}
\usepackage[T1]{fontenc}
\usepackage[latin9]{inputenc}
\usepackage{units}
\usepackage{textcomp}
\usepackage{amstext}
\usepackage{amsmath}
\usepackage{amssymb}
\usepackage{graphicx}
\usepackage{esint}
\usepackage{pifont}
\usepackage{adjustbox}
\usepackage{float}

\begin{document}
\providecommand{\tabularnewline}{\\}


\title{Characterizing the nonlinear interaction of S- and P-waves in a rock sample} 



\author{Thomas Gallot}
\affiliation{Massachusetts Institute of Technology, Cambridge, USA. Now at Science
Institute of Universidad de la Republica, Uruguay}

\author{Alison Malcolm}
\affiliation{Massachusetts Institute of Technology, Cambridge, USA.  Now at Memorial University of Newfoundland, Canada.}

\author{Thomas L. Szabo}
\affiliation{Boston University, Boston, USA}

\author{ Stephen Brown, Daniel Burns, Michael Fehler}
\affiliation{Massachusetts Institute of Technology, Cambridge, USA}

\begin{abstract}
The nonlinear elastic response of rocks is known to be caused by the
rocks' microstructure, particularly cracks and fluids. This paper
presents a method for characterizing the nonlinearity of rocks in
a laboratory scale experiment with a unique configuration. This configuration
has been designed to open up the possibility the nonlinear characterization
of rocks as an imaging tool in a field scenario. The nonlinear interaction
of two traveling waves: a low-amplitude 500~kHz P-wave probe and
a high-amplitude 50~kHz S-wave pump has been studied on a room-dry
15 x 15x 3 cm slab of Berea sandstone. Changes in the arrival time
of the P-wave probe as it passes through the perturbation created
by the traveling S-wave pump were recorded. Waveforms were time gated
to simulate a semi-infinite medium. The shear wave phase relative
to the P-wave probe signal was varied with resultant changes in the
P-wave probe arrival time of up to 100~ns, corresponding to a change
in elastic properties of $0.2\%$. In order to estimate the strain
in our sample, ae also measured the particle velocity at the sample
surface to scale a finite difference linear elastic simulation to
estimate the complex strain field in the sample, on the order of $10^{-6}$,
induced by the S-wave pump. We derived a fourth order elastic model
to relate the changes in elasticity to the pump strain components.
We recover quadratic and cubic nonlinear parameters: $\tilde{\beta}=-872$,
$\tilde{\delta}=-1.1\times10^{10}$, respectively, at room-temperature
and when particle motions of the pump and probe waves are aligned.
Temperature fluctuations are correlated to changes in the recovered
values of $\tilde{\beta}$ and $\tilde{\delta}$ and we find that
the nonlinear parameter changes when the particle motions are orthogonal.
No evidence of slow dynamics was seen in our measurements. The same
experimental configuration, when applied to Lucite and aluminum, produced
no measurable nonlinear effects. In summary, a method of selectively
determining the local nonlinear characteristics of rock quantitatively
has been demonstrated using traveling sound waves.
\end{abstract}

\maketitle 

\section{Introduction}

Mechanical waves provide information for characterizing the bulk properties
of materials noninvasively. Classical methods usually create a map
of linear information, such as elastic modulus, to detect structures.
Imaging structures is just a beginning; many applications require
more specific information with the goal of determining the quantitative
nature of the structures. In rocks, nonlinear elastic properties vary
over several orders of magnitude \cite{Belyayeva_1995} making them
good candidates for imaging. This nonlinearity is primarily due to
the microstructure of the rocks \cite{Guyer_1999,Guyer_2009}. An
understanding of this microstructure is increasingly important for
subsurface exploration. This study aims to characterize the nonlinearity
of rocks in a laboratory scale experiment with a configuration that
mimics a potential field scenario. In the experiment we perturb the
propagation of a low amplitude high frequency P-wave probe with a
high amplitude low frequency S-wave pump. We use a configuration with
a large distance between the probe source and receiver (30 probe wavelengths)
and a propagating pump wave. This experiment is designed as a preliminary
study working toward an imaging method based on the nonlinear interaction
of two waves.

Guyer \textit{et. al} \cite{Guyer_1999} demonstrated that nonlinearities
in rocks can be observed with strains as low as $10^{-8}$, this level
of sensitivity means that almost any kind of wave propagation can
induce a nonlinear effect; the challenge is in its detection. Field
observation of nonlinear responses induced by strong or weak earthquakes
are well documented (see \cite{Regnier_2013} for instance), and wave-speed
variations on the order of 0.05\% have been measured during earthquakes
on the San Andreas fault \cite{Brenguier_2008}. Actively induced
nonlinear responses have been observed \textit{in-situ} at the scale
of a few meters \cite{Kurtulus_2008,Johnson_2009,Lawrence_2009,Cox_2009}. 

Laboratory measurements are also helpful in understanding the nonlinear
elastic response. Of particular importance is the role of additional
compliance due to micro-cracks including anisotropy and fluid saturation
effects \cite{Holcomb_1981,Thomsen_1995,SAYERS_1995,Gueguen_2003,Fortin_2007}.
These studies were based on changes in acoustic signals under quasi-static
uni-axial stress or hydrostatic pressure. Because of the difficulty
in measuring the small changes induced by nonlinearities at small
strains ($10^{-8}-10^{-5}$), most laboratory studies of nonlinearity
in rocks use samples in resonance \cite{JOHNSON_1991,TenCate_1996,Guyer_1999,Gusev_1998,DAngelo_2004}.
At these strain amplitudes, no plastic deformation occurs and tiny
perturbations of soft bonds are responsible for the nonlinear behavior
\cite{Darling_2004}. These methods are generally based on monitoring
the frequency and damping of particular resonances and thus they average
the nonlinear response during several cycles of tension/compression.

To avoid this averaging, Dynamic Acousto-Elasticity (DAE) attempts
to account for the dynamics of the nonlinear interaction within a
cycle and is the closest method to the one proposed here. First developed
in medical characterization of bone and other materials \cite{Renaud_2008,Renaud_2009,Muller_2006},
it was later applied to rocks \cite{Renaud_2012,Renaud_2013,Renaud_2013a,Riviere_2013}.
The method relies on monitoring the material several times during
a resonant period. The monitoring is performed with a very low strain
probe wave and the high strain resonant wave is called the pump. This
probe wave monitors changes in the ultrasound properties, (wave-speed
and attenuation), during the quasi-static compression and tension
of the material, caused by the pump wave. This method takes advantage
of a uniform strain along a short probe path due to both a 1-D geometry
and the resonance of the sample. The ideas of DAE have also been applied
for \textit{in-situ} measurement by Renaud {\em et al.} \cite{Renaud_2013b}.

With the goal of developing an imaging technique for the nonlinear
elastic properties, we propose a DAE experiment with a unique configuration.
First, the probe source-receiver distance is large compared to the
pump wavelength. This allows us to estimate the nonlinearity not only
locally around a source-receiver pair, but also in a larger region.
Second, the resonant pump wave is replaced by a propagating wave,
time-gated to mimic an infinite sample. We then measure the change
in the arrival time of the probe as the pump wave crosses its path.
Finally, the P-wave pump is replaced with an S-wave allowing us to
change the relative particle motions of the pump and probe, by varying
the polarization of the shear wave. In the following we detail the
motivations for these three unique aspects of our experiment: large
source-receiver distance, propagating pump and S-wave pump.

The possibility of nonlinear parameter tomography for a large source-receiver
distance is treated theoretically for a harmonic field in Belyayeva
\textit{et. al} \cite{Belyayeva_1995}. Because we use a propagating
wave in our experiment, the strain is neither uniform nor static along
the probe path. The homogeneous strain distribution assumption also
does not hold in Geza {\em et al.} \cite{Geza_2001}, where an
attempt at nonlinear imaging is presented. 

The propagating pump wave is common to all \textit{in-situ} methods
\cite{Kurtulus_2008,Johnson_2009,Lawrence_2009,Cox_2009} and has
also been tested for a DAE method in \cite{Renaud_2013b}. In these
methods, the strain is generally estimated using embedded instruments.
At the laboratory scale, a different option is preferred. Our sample
mimics a 2D medium because the source wave transducer has a diameter
approximately the same size as the smallest dimension of the sample.
We can thus measure the strain on the surface and reasonably infer
its distribution within the sample with finite difference modeling,
resulting in an estimation of the strain distribution as a function
of time. The pump wave field is different than it would be in a semi-infinite
medium due to differences in geometrical spreading and conversions
at the surface. However, the pump wave remains a propagating wave
which is sufficient to illustrate the feasibility of the method for
\textit{in-situ} measurement. In addition, preliminary measurements
in a cube show similar results

Both nonlinear propagation of seismic waves and \textit{in-situ} methods
of nonlinear characterization involve shear strain components, the
interaction of which is the underlying physics of the nonlinear elasticity.
There is a gap between field observations and laboratory experiments
in rocks because, as far as the authors know, the nonlinear perturbation
of the medium is usually considered to be due to only compressive
strain components and does not typically include shear strain components.
The reason for this may come from the absence of quadratic nonlinearity
induced by a shear strain \cite{Belyayeva_1995}. Laboratory rock
experiments include nonlinear effects on shear wave propagation such
as shear wave splitting under uni-axial stress \cite{Toupin_1961}
and shear wave generation from P-wave mixing \cite{Johnson_1989},
but this does not include a shear strain component in the origin of
the nonlinearity. The choice of a shear wave pump in this paper aims
to consider a realistic pump strain field in a subsurface experiment,
that includes shear components. As for experiments, we found no theoretical
studies on the effect of shear strain on nonlinear elasticity; to
rectify this a fourth order elasticity model is presented, inspired
by a series of papers by Destrade \textit{et. al }\cite{Destrade_2010,Destrade_2010a,Destrade_2010b}.

The nonlinear characterization technique is presented in section~\ref{sec:I},
by discussing the experimental set-up, signal acquisition procedure,
and strain estimation method. A fourth order elasticity model is introduced
in section~\ref{sec:I}, followed by the definition of the nonlinear
parameters measured experimentally. Experimental results are then
presented in section~\ref{sec:Experimental-results} to characterize
the nonlinear response of a room-dry Berea Sandstone sample.

\section{Nonlinear wave mixing experiment\label{sec:I}}

The characterization of nonlinearity requires two fundamental measurements.
First the effect of the pump wave on the probe propagation is determined
from the modulation of the propagation time through the sample. And
second, the strain induced by the pump wave has to be measured in
order to quantify the nonlinearity. This second step is done by the
use of both a laser vibrometer to estimate the strain at the surface,
and also numerical modeling of the pump propagation to estimate the
strain in the whole sample.

\subsection{Experimental set-up\label{subsec 2.1}}

\begin{figure}[h]
\includegraphics[width=0.5\textwidth]{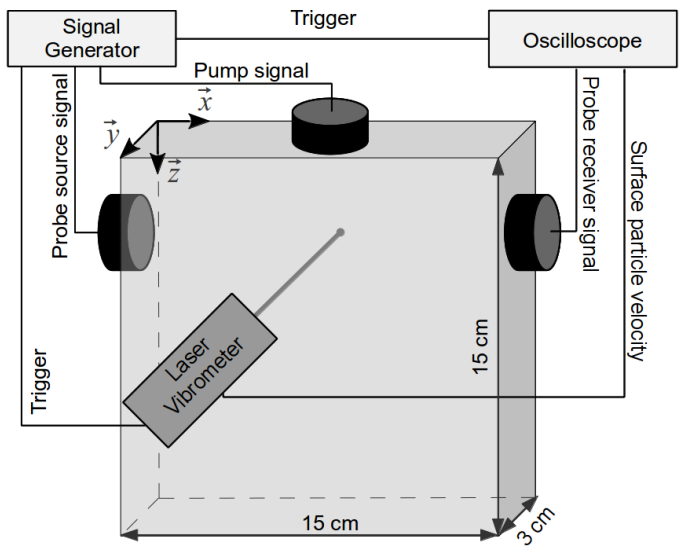} \protect\caption{\label{fig:1}Experimental set-up: a P-wave transducer generates an
ultrasonic pulse at 500 kHz in a $15\times15\times3$ cm sample. This
probe signal is recorded, after propagation through the sample, by
an identical transducer. The pump signal is generated with an S-wave
transducer at 50 kHz. The particle velocity of the pump, polarized
along the $x-$axis is measured with a laser vibrometer. The probe
signal and particle velocity are digitized at 250 MHz by an oscilloscope
triggered with the signal generator. The reference coordinate system
has a $+x$ axis along the probe propagation (left to right), $+z$
axis pointing down and the $+y$ axis perpendicular to the large surface.}
\end{figure}

Figure~\ref{fig:1} shows the experimental setup. We use a $15\times15\times3$~cm
block of Berea Sandstone, with linear properties summarized in Table~\ref{tab:1}.
We choose Berea Sandstone because it is relatively well-studied as
well as relatively homogeneous. We generate the low-amplitude (strain
less than $10^{-7}$, see section~\ref{sub:Quantifying-the-pump-strain})
500~kHz probe signal with a P-wave transducer with a 2.5~cm diameter
(Olympus Panametrics Videoscan V102-RB) on the left face of the sample
(i.e. propagating in the $+x-$direction); we record all signals with
an identical P-wave transducer on the opposite face of the sample.
The high-amplitude (strain around $10^{-6}$, see section~\ref{sub:Quantifying-the-pump-strain})
S-wave pump ($F_{0}=50$~kHz) is transmitted from a S-wave transducer
with a 2.5~cm diameter (Olympus Panametrics Videoscan V1548), placed
on the top face of the sample (i.e. propagating in the positive $z-$direction
with its particle motion aligned along the $+x$-axis in the $xz$-plane).
The method for estimating the strain is described below; for the probe
we had to amplify the signal so that it would be visible to the vibrometer,
and then linearly scale the estimated strain back to the levels used
in the original experiment. This gives us an order-of-magnitude estimate
of the probe strain of $10^{-7}$ during the experiment. Even at these
low strains, the probe wave is shown \cite{Riviere_2013} to have
an effect on the nonlinear response, but this effect is limited to
a slow dynamic effect (signals changing on the order of seconds to
hours), which is independent of the pump period \cite{Johnson_2005}.
Slow dynamic effects are not observed in this experiment as demonstrated
in section~\ref{sub:Memory-free-measurements}. The higher amplitude
S-wave pump does perturb the elastic properties of the medium, and
it is these perturbations that we are interested in measuring via
their nonlinear interaction with the P-wave probe. These interactions
remain small, however, and so we compare the probe signal with and
without the pump as described in section~\ref{sec:proc}. We record
three signals for each data point. The probe alone \ding{172}, the
pump and probe together \ding{173}, and then the pump alone \ding{174}.
Each signal is independently averaged by the scope 16 times, before
moving on to the next signal. Each signal is recorded for a duration
of 20~$\mu$s. The entire sequence \ding{172}\ding{173}\ding{174}$_{\phi_{0}}$
is recorded for a probe/pump delay $\phi=\phi_{0}$. Then the sequence
is repeated for different delays: \ding{172}\ding{173}\ding{174}$_{\phi_{0}}$,\ding{172}\ding{173}\ding{174}$_{\phi_{1}}$,\ding{172}\ding{173}\ding{174}$_{\phi_{2}}$...
We vary the delay between the probe and pump signal over several periods
of the pump to see the change in the probe traveltime as a function
of the phase of the pump.

\begin{table}
\begin{centering}
\begin{tabular}{r|r}
\hline 
Compressional wave speed  & $c_{p}=2450$ m/s \tabularnewline
\hline 
Shear wave speed  & $c_{s}=1550$ m/s \tabularnewline
\hline 
Density  & $\rho=2700$ kg/m\textthreesuperior{}\tabularnewline
\hline 
Elastic modulus $\lambda+2\mu$ & $M=16$ GPa\tabularnewline
\hline 
Length  & $L=15\:\mathrm{cm}$\tabularnewline
\hline 
High  & $H=15\:\mathrm{cm}$\tabularnewline
\hline 
Thickness  & $e=3\:\mathrm{cm}$\tabularnewline
\hline 
\end{tabular}
\par\end{centering}

\protect\caption{\label{tab:1}Berea Sandstone sample parameters}
\end{table}

We excite the probe wave at a much lower frequency than the pump so
that we can consider the pump wave to be in a steady-state during
the probe propagation. For our experiment, the ratio of the excited
P-wave probe wavelength to that of the S-wave pump is about 1/6, although
the recorded difference is somewhat smaller (see Figure~\ref{fig:2}),
due to dispersion in the sample. The choice of a shear wave for the
pump allows us to control the main direction of strain and gives a
slower change in the strain distribution. The number of cycles of
the pump signal is chosen to avoid reflections from the bottom of
the sample ($z=15$~cm) in order to mimic a semi-infinite medium
with no resonance. The wavelength at this frequency is $\lambda_{s}=3.1$~cm
so, with 6 cycles, and a return-time of 200~$\textrm{\ensuremath{\mu}s}$,
there is no reflection returning within the time of the probe propagation
across the sample (60~$\mu$s). The maximum delay of the probe excitation
(after the pump excitation) is $\phi=120~\mu$s. After probe excitation,
the total observation time (180~$\mu$s) is still less than the return
time. The probe wavelength ($\lambda_{p}=4.5$~mm) ensures that the
perturbation induced by the shear wave pump is uniform as seen by
the probe propagation. The phase delay between the probe and the pump
signals is changed to scan several cycles of tension and compression
induced by the pump.

We use an arbitrary waveform generator to create the probe, pump and
trigger signals. A power amplifier is needed for the pump signal in
order to reach sufficiently high strains. At the receiving P-wave
transducer, we are interested in the probe signal and not in the pump;
obviously when the pump and probe are both active, we record both
signals. To mitigate this, a second order high-pass frequency filter,
with a cut-off frequency $f_{c}=150$~kHz, is used to minimize the
amplitude of the pump signal measured at the receiver, so that mainly
the probe signal is recorded. The attenuation of the filtering is
compensated with a pre-amplifier (+50 dB). In addition, we use a low-pass
filter, cut-off frequency $f_{c}=1.5$~MHz to eliminate some high-frequency
noise. The acquisition of the probe signal by the P-wave receiver
and the shear pump displacement measured by the laser vibrometer are
synchronized via the trigger signal. The electronics are fully controlled
via MATLAB: transmission and receiving parameters, as well as the
recording of the signals. The delay $d_{p}=8$~ms between two consecutive
acquisitions, for example between \ding{172} and \ding{173} is chosen
to avoid the superposition of consecutive signals, i.e. to avoid recording
the reverberation of the shear wave pump in the sample. For the same
reason the delay is the same between two consecutive sequences \ding{172}\ding{173}\ding{174}$_{\phi_{0}}$
and \ding{172}\ding{173}\ding{174}$_{\phi_{1}}$.

\subsection{Nonlinear signal observation\label{sec:proc}}

\begin{figure}[h]
\includegraphics[width=0.5\textwidth]{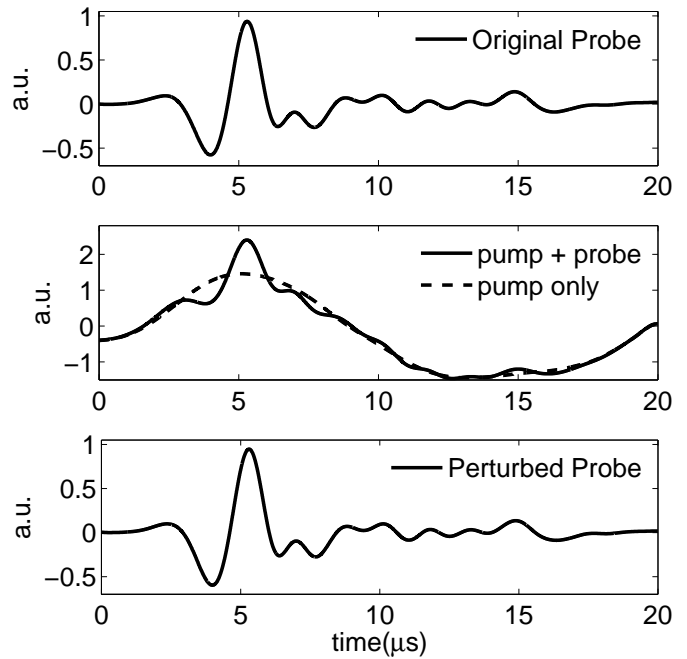}

\protect\caption{\label{fig:2}a) First, the original response to the probe pulse with
no perturbation (\ding{172} solid line) is recorded on the receiver.
b) Next, we turn on the shear wave pump to record the superposition
of the pump and probe signals (\ding{173} solid line). Finally, the
response to the pump with no probe is also recorded (\ding{174} dashed
line). c) The perturbed probe, constructed from the difference between
solid and dashed line in b), is compared to the original probe at
each phase. In this example the phase shift between pump and probe
is $\phi=25~\mu s$. Note that at 50 kHz, the pump signal is attenuated
by -54 dB with a high-pass filter. \textquotedblleft Arbitrary units\textquotedblright{}
is abbreviated as \textquotedblleft a.u.\textquotedblright{}}
\end{figure}

Each data point is obtained from the three signals shown in Figure~\ref{fig:2}.
First, we record \ding{172} the probe pulse with no pump present,
shown in a). Second, we record \ding{173} the perturbed probe with
the pump turned on: solid line in b). Third, we record \ding{174}
the pump signal alone (dashed line, b). We then subtract the pump
signal (dashed) from the perturbed probe and pump signal (solid) to
estimate the perturbed probe, shown in c). The perturbed probe, Figure~\ref{fig:2}
c), is compared to the original one to estimate the nonlinear perturbation.

The measured arrival time modulation $TM_{meas}$, induced by the
interaction over the propagation path of the probe wave with the pump
wave, is defined as 
\begin{eqnarray}
TM_{meas}\left(\phi\right) & = & T_{p}\left(\phi\right)-T_{o}\left(\phi\right),\label{Time modulation}
\end{eqnarray}
where $T_{o}$ is the arrival time of the original probe, $T_{p}$
that of the perturbed probe, and $\phi$ is the phase shift (a time
delay added to the transmitted pulse) between the probe and pump signals.
$TM$ is measured by cross-correlating the original (shown in Figure~\ref{fig:2}a))
and the perturbed (shown in Figure~\ref{fig:2}c)) pulses. The correlation
is computed in a two period window, centered on the maximum of the
signal ($3.3\,\mu s<t<7.3\,\mu s$ in Figure~\ref{fig:2}). The changes
in travel time are small, much smaller than the time sampling interval,
so we interpolate the peak of the cross-correlation with a second-order
polynomial before picking the maximum \cite{Catheline_1999}. We discard
data for which the waveforms change, defined as a correlation coefficient
of less than 0.99. We observed that the subtraction of the low frequency
part of the signal does not modify the waveform, and that the perturbation
is small enough to neglect any stretching of the probe pulse due to
distortion of the waveform.

\begin{figure}[h]
\includegraphics[width=0.5\textwidth]{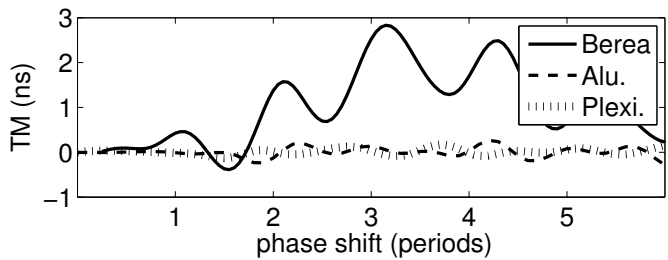} \protect\caption{\label{fig:3} A comparison of the nonlinear response of different
materials. The responses in both Lucite and Aluminum shown as dashed
and dotted lines are significantly smaller than those recorded for
Berea Sandstone as a solid line. }
\end{figure}

The $TM_{meas}$ between the original and perturbed probe is shown
as a function of the phase shift between the probe and pump in Figure~\ref{fig:3},
solid line). Each point is an average of 30 acquisitions. We apply
a low pass filter with a cutoff frequency at 100~kHz (twice the pump
frequency) to $TM_{meas}(\phi)$ to remove high frequency components
induced by noise. The signal $TM_{meas}(\phi)$ clearly has two frequency
components, one around the pump frequency as well as a very low frequency
trend. The presence of the pump frequency suggests that $TM_{meas}$
contains a term proportional to the pump strain, this is the so-called
quadratic nonlinearity. Then, the low frequency trend requires an
additional term that is always positive. The most likely candidate
is the square of the strain; this cubic linearity is known to be large
in rocks\cite{Guyer_1999}. Because the probe wave experiences approximately
three tension-compression cycles during its propagation from source
to receiver, the hysteresis known to play an important role rocks
(\cite{Holcomb_1981}) can not be clearly observed.

We repeated the experiment in aluminum and lucite, as shown in Figure~\ref{fig:3}
the measured time modulation is very small in these materials ($\pm0.2$~ns),
without any clear component at the pump frequency. These signals are
at least an order of magnitude higher than what we would expect for
aluminum ($TM\approx0.5$~ns for a $10^{-5}$ strain according to
\cite{Renaud_2012}; our strain is $\sim10^{-6}$) and are almost
certainly experimental noise. In Lucite, the nonlinear parameters
are even smaller (see \cite{Winkler_1996}), confirming the significance
of both aluminum and Lucite measurements. This comparison with standard
linear material ensures that $TM_{meas}$ does not originate in the
lab equipment, but depends on the sample studied. 

\subsection{Estimating the strain\label{sub:Quantifying-the-pump-strain}}

As will be shown in the following section, the characterization of
the nonlinearity directly relies on the estimation of the strain pump.
We thus need to characterize the pump field within the sample. At
the order of magnitude ($<10^{-5}$ in strain) and at the pump frequency
(50 kHz) we are considering, direct measurements are impossible to
perform with strain gauges. Previous studies used a laser vibrometer
to measure the particle velocity at a particular point on the sample,
and then interpolated the strain assuming vibration at a single resonance
\cite{Renaud_2012}. A similar method is used in this experiment,
but since the wave-field is not a single resonance, we require a more
careful numerical modeling of the wavefield.

\subsubsection{Numerical model}

We use a 3D, isotropic, purely-elastic (i.e. lossless) finite difference
model, following the method described in references \cite{Virieux_1986,Graves_1996},
to model the linear propagation of the shear wave pump. We apply free-surface
boundary conditions on all sides of the sample, and compute the stress
and particle velocities on a staggered grid. The same geometry and
wave speeds mentioned in section~\ref{subsec 2.1} are used as input
parameters to the model; the geometry is shown in figure~\ref{fig:1}.
The spatial meshing of the model is 0.5~mm and the time sampling
is 0.08~$\textrm{\textrm{\ensuremath{\mu}}s}$.

The challenge in modeling this experiment is in obtaining an accurate
model of the transducer, so that the modeled and recorded waveforms
match one another. We model the transducer with 1250 point force sources
distributed on a disk with a diameter of 2.5 cm. We then use the x-component
of the particle velocity ($V_{x}$) recorded by a laser vibrometer
(polytech CLV-3D, see Fig. \ref{fig:1}) as the input force signal
for the simulation. In other words, we record the particle velocity
experimentally, and then use that signal to drive the simulated transducer.
Note that the laser signal records the surface particle velocity at
the position ($x=7.5,\: y=3,\: z=3.6$~cm) while the shear transducer
creates a force along the $x-$axis at the position ($x=7.5,\: y=1.5,\: z=0$~cm).
Because of this, we scale the amplitude of the numerical result to
match that of the experiment. This is valid because we are doing a
linear simulation. The scaling of the model using only a single point
measurement of particle velocity may induce errors in the strain estimation,
particularly when there are diffraction effects.

\subsubsection{Pump strain field}

\begin{figure}[h]
\includegraphics[width=0.5\textwidth]{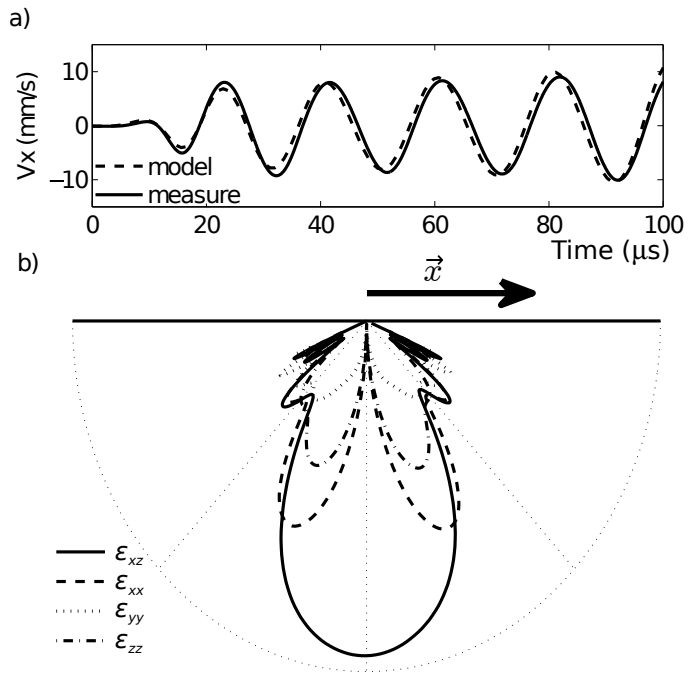}\protect\caption{\label{fig:4} a) The x-component of the particle velocity at ($x=7.5,\: y=3,\: z=3.6$
cm) is measured with a laser vibrometer (solid line) and modeled with
a finite difference simulation (dashed line). b) Polar radiation patterns
of the shear transducer are shown for the shear strain $\epsilon_{xz}$
(solid line), and the compressive strains $\epsilon_{xx}$ (dashed
line), $\epsilon_{yy}$ (dotted line), and $\epsilon_{zz}$ (dash-dotted
line). The arrow represents the transducer force direction ($x-$axis).}
\end{figure}
Figure~\ref{fig:4} a) shows good first order agreement between $V_{x}$
measured by the laser and modeled, after calibration. The apparent
small difference in frequency between the two signals could be caused
by a number of things, the most likely of which is interferences of
different wave types. From the calibrated simulation, we obtain the
stress throughout the sample, at all times. We then compute the strain,
using a linear Hooke's law. 

The elastic response of the sample does not contain a pure shear wave
since the transducer has a finite size. The radiation pattern of the
transducer is represented in polar coordinates in Figure~\ref{fig:4}
b) for different strain components. The radiation pattern represents
the relative amplitude of each strain as a function of the angle in
the $xz$-plane for a strain $\epsilon_{ij}$ and at a 3-cm distance
from the transducer. The main strain component is the shear $\epsilon_{xz}$
strain that corresponds to the propagating S-wave pump. The strain
magnitudes can be compared by computing the absolute maxima of each
strain ($\Vert\epsilon_{ij}\Vert$). The following ratios are found
:
\begin{equation}
\Vert\epsilon_{xz}\Vert\approx1.5\Vert\epsilon_{xx}\Vert\approx2\Vert\epsilon_{zz}\Vert\approx3\Vert\epsilon_{yy}\Vert\approx20\Vert\epsilon_{xy}\Vert\approx40\Vert\epsilon_{yz}\Vert.\label{relative strain amplitude}
\end{equation}
The compressive strains are on the same order of magnitude but the
shear strain $\epsilon_{xy}$ and $\epsilon_{yz}$ can be neglected
and thus are not represented in Figure~\ref{fig:4}. The components
$\epsilon_{xz}$ and $\epsilon_{xx}$ have a similar pattern to that
of the tangential and radial component of the displacement field created
by a shear transducer in an elastic half-space. The other two components
$\epsilon_{yy}$ and $\epsilon_{zz}$ are not present in a half-space
but arise from the limited size of the sample along the $y-$direction
and thus need to be taken into account in the nonlinear characterization
of the material.

\subsubsection{Probe strain field}

For the probe strain estimation we apply a similar method with a few
changes. First, another laser vibrometer was utilized to achieve a
higher sensitivity around the probe frequency 500kHz (a Polytech system
with a OFV-505 optics head controlled by OFV-5000 with VD-06 decoder).
Second, a sufficient amplitude of excitation was used to obtain a
signal significantly above the noise. The particle velocity was deduced
by increasing the input amplitude and then scaling the laser-measured
amplitude. The input maximum voltage for the probe source transducer
was set at 10 times the usual voltage: 20~V instead of 2~V. With
a 2-V input signal, only the transducer was sensitive enough to measure
a signal; the laser vibrometer signal was too noisy to obtain a reliable
signal. The linearity of the transducer emission was checked by comparing
the acoustic signal recorded by the probe receiver (Figure~\ref{fig:1})
with a 2-V and a 20-V input. Both signals have the same waveform and
vary by a factor 10 in amplitude. We measured the resulting $x-$component
of the particle velocity close to the probe receiver at ($x=15,\: y=1.5,\: z=1$~cm).
We then divide the particle velocity by a factor 10 to scale the amplitude
of the numerically modeled strain data. In addition to this scaling,
the positions and the directions of the point force sources were changed
for the probe simulation (see Figure~\ref{fig:1}). The result shows
that the volumetric strain decreases nearly linearly along the propagation
path from $9\times10^{-8}$ around $x=0$~cm to $4\times10^{-8}$
around $x=15$~cm. The value of $10^{-7}$ used above can thus be
thought of as a rough upper bound for the strain excited by the probe.

\section{Theoretical description of the nonlinearity}

In this section we establish a fourth order nonlinear Hooke's law
that relates the pump strain to the elasticity variation. This model
depends on many elastic moduli that can not all be measured in the
present experiment. We then present an approximation of the model
and two nonlinear coefficients are defined. Finally we relate the
measured change in travel-time to the nonlinear coefficients.

\subsection{Fourth order elasticity theory \label{FOET}}

As mentioned in section~\ref{sec:proc}, the present experiment requires
a model of the elastic response which includes both quadratic and
cubic nonlinearities. In addition, the probe wave interacts with the
pump wave over several cycles and we observe no hysteresis. Consequently,
hysteresis is not included in the model. The description of a nonlinear
elastic system starts from the strain energy $W$. The stress $\sigma_{ij}$
associated with the strain is then given by 
\begin{equation}
\sigma_{ij}=\frac{\partial W}{\partial\epsilon_{ij}}\,,\label{stress-strain energy}
\end{equation}
where $\epsilon_{ij}=\frac{\partial u_{i}}{\partial x_{j}}$ is the
Eulerian strain with $u_{i}$ the displacement along the $i$-axis
with $i=x,y,z$. The derivative in Eq.~\ref{stress-strain energy}
implies that the strain energy must be fourth-order in the strain
to result in a third order (cubic) stress-strain relationship. For
an isotropic material (we neglect the4\% anisotropy measured in our
sample) Landau and Lifshitz \cite{Landau_1986} show that the strain
energy can be described by the three invariants: 
\begin{equation}
I_{k}=tr(L^{k}),\qquad k=1,2,3\label{eq:invariants-1}
\end{equation}
where $L$ is the Lagrangian strain : $L_{ij}=\frac{1}{2}(\epsilon_{ij}+\epsilon_{ji}+\sum_{k}\epsilon_{ki}\epsilon_{kj})$.
The subscript denotes the minimum order of the invariant $I_{1,2,3}$;
these invariants are the traces of $L$, $L^{2}$, $L^{3}$, respectively.
For linear elasticity only the first two orders are considered: $I_{1}$
and $I_{2}$. Landau \& Lifshitz \cite{Landau_1986} write the strain
energy $W_{3}$ up to third order by including terms in $I_{3}$,
$I_{1}I_{2}$ and $I_{1}^{3}$. There are 4 combinations of the invariants
in the strain energy at the fourth order: $I_{1}I_{3}$, $I_{1}^{2}I_{2}$,
$I_{2}^{2}$ and $I_{1}^{4}$, thus the fourth-order strain energy
is \cite{Zabolotskaya_1986,Jacob_2007,Destrade_2010b,Abiza_2012}:
\begin{equation}
W=\frac{\lambda}{2}I_{1}^{2}+\mu I_{2}+\frac{A}{3}I_{3}+BI_{1}I_{2}+\frac{C}{3}I{}_{1}^{3}+EI_{1}I_{3}+FI_{1}^{2}I_{2}+GI_{2}^{2}+HI{}_{1}^{4}\,,\label{Fourth order strain energy}
\end{equation}
where $A$, $B$ and $C$ are the third order elastic moduli introduced
by Landau-Lifshitz \cite{Landau_1986}, and $E,\; F,\; G$ and $H$
are the fourth order elastic moduli \cite{Zabolotskaya_1986}.

In order to understand the present experiment, we first consider the
ideal case of a P-wave probe propagating along the $x-$axis and a
pure S-wave pump propagating along $z-$axis polarized along the $x-$axis.
The probe and pump waves induce $\epsilon_{xx}$ and $\epsilon_{xz}$
strain components, respectively. The strain energy $W$, computed
as a function of these two strain components including terms at the
fourth order and below is 
\begin{equation}
W=\frac{M}{2}\epsilon_{xx}^{2}+\frac{\mu}{2}\epsilon_{xz}^{2}+\gamma_{1}\epsilon_{xx}^{3}+\gamma_{2}\epsilon_{xx}\epsilon_{xz}^{2}+\gamma_{3}\epsilon_{xx}^{4}+\gamma_{4}\epsilon_{xx}^{2}\epsilon_{xz}^{2}+\gamma_{5}\epsilon_{xz}^{4}\,.\label{W fourth experiment case}
\end{equation}
In Eq.~\ref{W fourth experiment case}, the linear elastic modulus,
$M$ is given by 
\[
M=\lambda+2\mu\,,
\]
the third order coefficients are 
\[
\gamma_{1}=\frac{M}{2}+\frac{A+3B+C}{3}\,,
\]

\[
\gamma_{2}=\frac{M}{2}+\frac{A+2B}{4}\,,
\]
and the fourth order coefficients are 
\[
\gamma_{3}=\frac{M}{8}+\frac{A+3B+C}{2}+E+F+G+H\,,
\]
\[
\gamma_{4}=\frac{M}{4}+\frac{5A+14B+4C}{8}+\frac{3E+2F+4G}{4}\,,
\]
and 
\[
\gamma_{5}=\frac{M}{8}+\frac{A+2B}{8}+\frac{G}{4}\,.
\]
The stress is computed from the strain energy by $\sigma_{ij}=\frac{\partial W}{\partial\epsilon_{ij}}$;
because we are interested in changes in the probe wave, we require
only $\sigma_{xx}$, the probe stress 
\begin{equation}
\sigma_{xx}=M\epsilon_{xx}+3\gamma_{1}\epsilon_{xx}^{2}+\gamma_{2}\epsilon_{xz}^{2}+4\gamma_{3}\epsilon_{xx}^{3}+2\gamma_{4}\epsilon_{xx}\epsilon_{xz}^{2}\,.\label{Stress in experimental case}
\end{equation}
In Eq.~\ref{Stress in experimental case}, the first term (Hooke's
law) is responsible for linear probe wave propagation, the second
and fourth terms are the quadratic and cubic nonlinearities in the
probe propagation respectively, and the third term governs the nonlinear
propagation of the pump. It is the fifth term $2\gamma_{4}\epsilon_{xx}\epsilon_{xz}^{2}$
that describes the interaction of the two waves. Renaming the probe
strain $\epsilon_{p}=\epsilon_{xx}$ to highlight the amplitude difference
between probe and pump: $\epsilon_{p}\ll\epsilon_{xz}$, we observe
that this interaction term is clearly the dominant nonlinear effect.
We then simplify Eq.~\ref{Stress in experimental case} to include
only the linear propagation and the interaction of the pump and probe
\begin{equation}
\sigma_{xx}=\epsilon_{p}\left(M+2\gamma_{4}\epsilon_{xz}^{2}\right)\,.\label{stress compressive wave}
\end{equation}
From Eq.~\ref{stress compressive wave}, the importance of the cubic
term in $\epsilon_{p}\epsilon_{xz}^{2}$ for the nonlinear coupling
is highlighted. In this case there is no quadratic coupling term in
($\epsilon_{p}\epsilon_{xz}$) because the corresponding term in strain
energy ($\epsilon_{p}^{2}\epsilon_{xz}$) is not present. Other pump
strain components will introduce this dependence. Including this stress
in the dynamic response of the elastic system gives a nonlinear (wave-like)
equation of propagation for the P-wave probe 
\begin{equation}
\rho\ddot{u}_{x}=\frac{\partial\sigma_{xx}}{\partial x}=\frac{\partial^{2}u_{x}}{\partial x^{2}}M\left(1+\frac{\mathrm{d}M}{M}\right)\,.\label{eq: fourth_order_elasticity}
\end{equation}
In section~\ref{sec:TMtoelastic}, we show that the nonlinear term
$dM/M$ is directly related to the measured arrival time modulation.
In the simplified example discussed here, $dM/M$ contains only a
cubic term: $dM/M=\delta_{xz}\epsilon_{p}\epsilon_{xz}^{2}$, with
$\delta_{xz}=2\gamma_{4}$ the cubic coefficient reported in line
3 of Table \ref{Table 2}.

\begin{table*}
\begin{adjustbox}{width=0.9\textwidth,center} %
\begin{tabular}{c|c|c}
pump strain ($\epsilon_{ij}$)  & quadratic coefficient ($\beta_{ij}$)  & cubic coefficient($\delta_{ij}$) \tabularnewline
\hline 
$\epsilon_{xx}$  & $3+2(A+3B+C)/M$  & $3/2+6\left(A+3B+C\right)/M+12\left(E+F+G+H\right)/M$\tabularnewline
\hline 
$\epsilon_{yy},\:\epsilon_{zz}$  & $\lambda/M+2(B+C)/M$  & $\lambda/2M+2\left(B+C\right)/M+4\left(F+G+3H\right)/M$\tabularnewline
\hline 
$\epsilon_{xz},\,\epsilon_{zx},\,\epsilon_{xy},\,\epsilon_{yx}$  & $0$  & $1/2+\left(5A+14B+4C\right)/4M+\left(3E+2F+4G\right)/2M$\tabularnewline
\hline 
$\epsilon_{yz},\:\epsilon_{zy}$  & $0$  & $\lambda/2M+\left(3B+2C\right)/2M+\left(F+2G\right)/M$\tabularnewline
\end{tabular}\end{adjustbox}\protect\protect\caption{\label{Table 2}Fourth order nonlinear coupling between a P-wave inducing
a compressive strain along the $x-$axis $\epsilon_{p}$ and all possible
pump strain components. In the stress-strain relationship, the quadratic
coefficient $\beta_{ij}$ weights the term in $\epsilon_{p}\epsilon_{ij}$,
and the cubic one $\delta_{ij}$ the term in $\epsilon_{p}\epsilon_{ij}^{2}$.
On the first line, $\epsilon_{xx}$ is the pump strain and do not
includes the probe strain $\epsilon_{p}$.}
\end{table*}

This ideal case of a pure shear wave illustrates the computation of
the fourth order wave mixing coefficients, but in Section~\ref{sub:Quantifying-the-pump-strain}
we note that the pump wave field is more complex than a pure shear
strain. We thus need to consider other strain components. For a P-wave
probe propagating along the $x-$axis, there are 4 combinations of
pump strain summarized in Table \ref{Table 2}. For each case the
strain energy is computed and the nonlinear stress-strain relationship
is obtained by differentiating $W$ with respect to $\epsilon_{xx}$
(as in Eqs. \ref{W fourth   experiment case} to \ref{stress compressive wave}).
The linear term remains unchanged since it relates to the linear propagation
of the probe and not to nonlinear wave mixing. Each combination of
strain gives one quadratic coefficient, which weights the coupling
between the probe strain $\epsilon_{p}$ and the pump strain $\epsilon_{ij}$,
and one cubic coefficient for the coupling with $\epsilon_{ij}^{2}$.
In the case of a P-wave pump along the $x-$axis, $\epsilon_{xx}$
includes both the probe strain $\epsilon_{p}$ and the pump strain
$\epsilon_{xx}$ . Substituting $\epsilon_{xx}\equiv\epsilon_{p}+\epsilon_{xx}$
along with $\epsilon_{xz}=0$ in Eq.~\ref{Stress in experimental case}
and neglecting the nonlinear propagation of waves ($\epsilon_{p}^{2}$,
$\epsilon_{p}^{3}$, $\epsilon_{xx}^{2}$ and $\epsilon_{xx}^{3}$)
yields the probe stress $\sigma_{xx}$

\begin{equation}
\sigma_{xx}=M\epsilon_{p}+6\gamma_{1}\epsilon_{p}\epsilon_{xx}+12\gamma_{3}\epsilon_{p}\epsilon_{xx}^{2}\,.\label{Stress in case P-P}
\end{equation}
The quadratic coefficient $\beta_{xx}=6\gamma_{1}/M$ and the cubic
coefficient $\delta_{xx}=12\gamma_{3}/M$ are reported in the first
line of Table \ref{Table 2}. The quadratic coefficients gathered
in Table \ref{Table 2} (second column) were obtained by Guyer \textit{et.
al} \cite[p. 47]{Guyer_2009} and the cubic coefficients are given
in \cite[p.268]{Hamilton_1998} as a function of the elastic tensors
($C_{ijk}$ and $C_{ijkl}$). The quadratic coefficient is nonzero
only with a compressive strain pump, this is confirmed in \cite[p. 266]{Hamilton_1998}
where the third order elastic tensor $M_{ijklmn}$ is found to be
zero for a shear strain pump ($i=j=k=l$, and $m\neq n$).

\subsection{Nonlinear coefficients approximation\label{sub:Nonlinear-coefficients-approxima}}

From the strain pump characterization, in Section~\ref{sub:Quantifying-the-pump-strain},
we know that $\epsilon_{xx},\:\epsilon_{yy},\:\epsilon_{zz}$ and
$\epsilon_{xz}$ are of the same order of magnitude; the other strain
components are at least an order of magnitude smaller and can thus
be neglected. Consequently a complete expression of the fourth-order
nonlinear elastic model of our experiment is given by 
\begin{equation}
\frac{\mathrm{d}M}{M}=\beta_{xx}\epsilon_{xx}+\beta_{yy}\epsilon_{yy}+\beta_{zz}\epsilon_{zz}+\delta_{xx}\epsilon_{xx}^{2}+\delta_{yy}\epsilon_{yy}^{2}+\delta_{zz}\epsilon_{zz}^{2}+\delta_{xz}\epsilon_{xz}^{2}\,.\label{full
  dm}
\end{equation}
This expression contains seven interconnected parameters, which include
all of the seven fourth-order elastic parameters. The experiment does
not allow us to estimate all parameters independently, we thus need
to simplify Eq.~\ref{full dm}. First of all, the order of magnitude
for linear, quadratic, and cubic elastic moduli are different (i.e.
$\lambda,\:\mu<<A,\: B,\: C<<E,\: F,\: G,\: H$) and we can thus neglect
terms containing linear moduli in the expressions for the quadratic
moduli, and terms with the linear and quadratic moduli in the expression
for the cubic moduli. Since only the quadratic and cubic nonlinearities
can be measured independently in our experiment, we need to go from
7 unknowns to only 2. One simple way to achieve this is to assume
that quadratic coefficients are of the same order of magnitude: $A\approx B\approx C$,
and the same assumption for the cubic coefficients: $E\approx F\approx G\approx H$.
This lead to a proportionality between the different coefficients
of the same order: $\frac{\beta_{xx}}{10}\approx\frac{\beta_{yy}}{4}\approx\frac{\beta_{zz}}{4}$
and \textbf{$\frac{\delta_{xx}}{48}\approx\frac{\delta_{yy}}{20}\approx\frac{\delta_{zz}}{20}\approx\frac{2\delta_{xz}}{9}$}.
Under these assumptions, the approximate nonlinearity is: 
\begin{align}
\frac{\mathrm{d}M}{M} &\approx\tilde{\beta}\left(\epsilon_{xx}+\frac{2}{5}\epsilon_{yy}+\frac{2}{5}\epsilon_{zz}\right)  \nonumber \\
&+\tilde{\delta}\left(\epsilon_{xx}^{2}+\frac{5}{12}\epsilon_{yy}^{2}+\frac{5}{12}\epsilon_{zz}^{2}+\frac{3}{32}\epsilon_{xz}^{2}\right)\,.\label{aprox dm-1}
\end{align}
The nonlinear parameters $\tilde{\beta}$ and $\tilde{\delta}$ are
coefficients of the quadratic and cubic nonlinearity respectively
and can be thought of as averaged elastic moduli : $\tilde{\beta}\equiv\frac{\left(A+B+C\right)}{3M}$,
$\tilde{\delta}\equiv\frac{\left(E+F+G+H\right)}{4M}$. They are representative
of the nonlinearity but can vary with the strain distribution since
the approximation implies that all strain invariants of the same order
play the same role in the strain energy (Eq.~\ref{Fourth order strain energy}).
Those parameters can also be considered as empirically defined since
only one parameter per order of nonlinearity can be measured with
one configuration of probe and pump waves.

Such a model is useful for describing the elastic response of the
rock at a fixed pump amplitude. Nevertheless, it does not capture
all the complexity of the mechanical response of the rock because
the nonlinear coefficients change with the pump amplitude as will
be shown in section~\ref{sub:Strain-amplitude-dependency}. The nonlinear
characterization of rocks depends on the amplitude of the perturbation,
this is why the quantification of the strain is so important. This
also implies that monitoring or imaging nonlinearities has to be done
with a constant pump amplitude to ensure repeatability.

\subsection{Relating measurements to the nonlinear parameters \label{sec:TMtoelastic}}

In a linear elastic medium the wave speed $c_{p}$ is constant or
equivalently the stress $\sigma$ is proportional to the strain $\epsilon$
as in Hooke's law: $\sigma=\epsilon M$, with the elastic modulus
$M=\lambda+2\mu$, where $\lambda$ and $\mu$ are the Lamé parameters.
A consequence of Hooke's law is a constant wave speed $c_{p}^{2}=M\rho^{-1}$,
with $\rho$ the density of the material. In this section, we detail
the necessary extension of Hooke's law for the nonlinear wave mixing
considered in this experiment. The arrival time modulation induced
by the pump strain (Fig. \ref{fig:3}), can be explained as a variation
of the wave speed $c_{p}$, or equivalently the elastic modulus $M$.
Assuming a homogeneous medium, $c_{p}$ can also be defined as $c_{p}=\ell_{x}T^{-1}$
where $T$ is the time of propagation along a distance $\ell_{x}$
(assumed to be along the $+x$-axis). Differentiating both expressions
for $c_{p}$ gives 
\begin{equation}
\frac{\mathrm{d}c_{p}}{c_{p}}=\frac{\mathrm{d}\ell_{x}}{\ell_{x}}-\frac{\mathrm{d}T}{T},\label{dcp dT}
\end{equation}
and 
\begin{equation}
\frac{\mathrm{d}c_{p}}{c_{p}}=\frac{\mathrm{d}M}{2M}-\frac{\mathrm{d}\rho}{2\rho}\,.\label{dcp dK}
\end{equation}
In rocks, the variation in distance of propagation $\mathrm{d}\ell_{x}$
and density $\mathrm{d}\rho$ can be neglected \cite{Renaud_2011}
in Eq.~\ref{dcp dT} and \ref{dcp dK} respectively. Equating these
two expressions shows that changes in time and elastic modulus are
proportional to one another, i.e. that 
\[
\frac{\mathrm{d}T}{T}=-\frac{\mathrm{d}M}{2M}\,.
\]
For an infinitesimal distance along the propagation path $dl$, the
propagation time is $T=dl/c_{p}$, the change in arrival time is then
\begin{equation}
\mathrm{d}T\approx-\frac{dl}{2c_{p}}\frac{\mathrm{d}M}{M}\,.\label{TOFM Moduli}
\end{equation}
To go from the infinitesimal changes in time and modulus to the observed
changes in travel time, we need to integrate equation~\ref{TOFM Moduli}
over the path length $\Gamma$ 
\begin{equation}
TM_{NL}=-\frac{1}{2c_{p}}\int_{\Gamma}\frac{\mathrm{d}M}{M}\, dl\,.\label{generic TMNL}
\end{equation}
In equation \ref{generic TMNL}, $TM_{NL}$ models the arrival time
modulation measured in Figure~\ref{fig:3} as a function of the variation
of the elastic nonlinearity $dM/M$, integrated over the propagation
path. In addition, the nonlinearity is a function of the pump strain
as defined in Eq.~\ref{aprox dm-1}. To estimate $TM_{NL}$ from
the pump strain as described in section~\ref{sub:Quantifying-the-pump-strain},
and because the pump transducer is approximately the same size as
the S-wave pump wavelength, the strain needs to be averaged within
the ultrasonic beam. For the sake of clarity, this average is included
in the strain notation: $\epsilon_{ij}\left(l,T\right)\equiv<\epsilon_{ij}\left(l,r,T)\right>_{r}$,
where $r$ is the radius of the beam. With this, along with the insertion
of $T=l/c_{p}$ into the time variable of the strain, the time modulation
from the nonlinear elasticity $TM_{NL}$ along the whole propagation
path becomes 
\begin{align}
TM_{NL}\left(\phi\right)=-&\frac{\tilde{\beta}}{2c_{p}}\int_{\Gamma}\epsilon\left(l,\phi+\frac{l}{c_{p}}\right)dl \nonumber\\
&-\frac{\tilde{\delta}}{2c_{p}}\int_{\Gamma}\epsilon^{2}\left(l,\phi+\frac{l}{c_{p}}\right)dl\,,\label{TOFM Moduli-1}
\end{align}
where $\epsilon=\epsilon_{xx}+\frac{2}{5}\epsilon_{yy}+\frac{2}{5}\epsilon_{zz}$
and $\epsilon^{2}=\epsilon_{xx}^{2}+\frac{5}{12}\epsilon_{yy}^{2}+\frac{5}{12}\epsilon_{zz}^{2}+\frac{3}{32}\epsilon_{xz}^{2}$,
these strains includes the average within the ultrasonic beam. In
the present experiment, with a homogeneous material, the propagation
path $\Gamma$ is a straight line of length $L$ along the $x-$axis.
In this case, the total arrival time modulation can be written as
\begin{equation}
TM_{NL}\left(\phi\right)=\tilde{\beta}\: Q\left(\phi\right)+\tilde{\delta}\: C\left(\phi\right)\,,\label{Total modulation-1}
\end{equation}
where the quadratic term is defined as: 
\begin{equation}
Q\left(\phi\right)=-\frac{1}{2c_{p}}\int_{0}^{L}\epsilon\left(x,\phi+\frac{x}{c_{p}}\right)dx\,,\label{eq:Q}
\end{equation}
and the cubic part is 
\begin{equation}
C\left(\phi\right)=-\frac{1}{2c_{p}}\int_{0}^{L}\epsilon^{2}\left(x,\phi+\frac{x}{c_{p}}\right)dx\,.\label{eq:C}
\end{equation}
These expressions state that the arrival time modulation can be computed
from the strains within the medium estimated as described in section~\ref{sub:Quantifying-the-pump-strain}.

\section{Experimental results\label{sec:Experimental-results}}

\subsection{Estimating the nonlinear parameters}

\begin{figure}[h]
\includegraphics[width=0.5\textwidth]{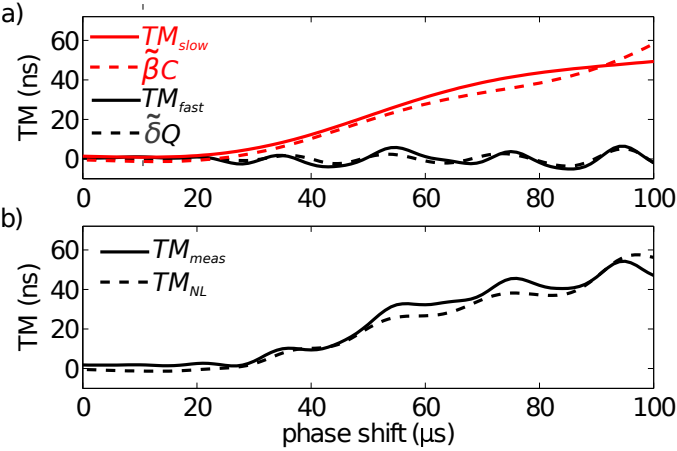} \protect\caption{\label{Fig 5}a) The nonlinear parameters $\tilde{\beta}=-872$, $\tilde{\delta}=-1.1\times10^{10}$
are computed from the fit between $TM_{fast}$ (black line) and $\tilde{\beta}\: Q\left(\phi\right)$
(dashed black) for the fast part, and between $TM_{slow}$ (red line)
and $\tilde{\delta}\: C\left(\phi\right)$ (dashed red) for the slow
part using Eq.~\ref{eq:delta_aprox} and \ref{eq:beta_aprox}. b)
The Time Modulation $TM$ computed from Eq.~\ref{Total modulation-1}
(dashed line) is in agreement with the measured signal $TM_{meas}$
(solid line).}
\end{figure}

The nonlinear parameters $\tilde{\beta}$ and $\tilde{\delta}$ are
then estimated by minimizing the difference between the arrival time
modulation computed from the modeled strain and particle velocity
(Eq.~\ref{Total modulation-1}) and the experimentally measured one
(Eq.~\ref{Time modulation}). It is helpful to point out that the
quadratic and cubic parts of the time modulation have different frequency
contents. If the pump signal is a monochromatic signal, which is close
to the observation of figure~\ref{fig:3}, then the strain can be
written as $\epsilon\propto\mathrm{cos}\left(\omega t\right)$, consequently
$Q\left(\phi\right)\propto\mathrm{cos}\left(\omega t\right)$ and
$C\left(\phi\right)\propto1+\mathrm{cos}\left(2\omega t\right)$.
This means that the two nonlinear parameters can be estimated separately
by frequency filtering around their corresponding dominant frequencies
via, 
\begin{equation}
TM_{meas}=TM_{slow}+TM_{fast}\,,\label{eq:slow_fast}
\end{equation}
where $TM_{slow}$ contains the frequencies down to $\frac{F_{0}}{2}$
($F_{0}=50$ kHz is the pump frequency) and $TM_{fast}$ includes
the frequencies between $\frac{F_{0}}{2}$ and $2\, F_{0}$. Then
$\tilde{\beta}$ and $\tilde{\delta}$ are computed with: 
\begin{eqnarray}
\tilde{\beta} & \approx & \frac{TM_{fast}}{Q}\,,\label{eq:beta_aprox}
\end{eqnarray}
\begin{eqnarray}
\tilde{\delta} & \thickapprox & \frac{TM_{slow}}{C}\,.\label{eq:delta_aprox}
\end{eqnarray}
These expressions make sense only for a perfect fit between measurement
and simulation where the ratios in Eq.~\ref{eq:delta_aprox} and
\ref{eq:beta_aprox} are constant for different values of the phase
shift $\phi$. Because of experimental error, modeling inaccuracies
etc, an error minimization is performed to estimate $\tilde{\beta}$
and $\tilde{\delta}$. Finally, the measured time modulation $TM_{meas}$
and the time modulation $TM$ computed from the strain as described
in Eq.~\ref{Total modulation-1} show good agreement as shown in
Figures \ref{Fig 5}. The travel-time perturbation begins after 20~$\textrm{\ensuremath{\mu}s}$,
which is when the pump wave reaches the probe's propagation path.
The pump oscillation induces the $TM_{fast}$ and slowly, the $TM_{slow}$
signal increases when the pump wave penetrates more and more of the
probe's path. The phase of the fast part of the signal $TM_{fast}$
is related to the quadratic nonlinearity $Q\tilde{\beta}$ and agrees
particularly well. The poorer agreement between $TM_{slow}$ and the
cubic nonlinearity and $C\tilde{\delta}$ is likely directly related
to the accumulation of errors between the experimental strain and
the estimated strain. It could also be a result of slow-dynamics,
although the discussion in section~\ref{sub:Memory-free-measurements}
indicates that this is less likely. The nonlinear quadratic parameter
found in this experiment: $\tilde{\beta}=-872$ is the same range
as the quadratic nonlinearity determined for 1-D nonlinear elastic
model with similar materials \cite{Renaud_2012,Riviere_2013} (where
the pump induces only $\epsilon_{xx}$ strain component). We found
$\tilde{\delta}=-1.1\times10^{10}$ for the cubic parameter which
is around two orders of magnitude bigger than in similar studies.
The nonlinear parameters are defined differently in those studies,
however discrepancies of about an order of magnitude have been observed
for different samples of the same type of rocks (Lavoux Limestone,
\cite{Renaud_2011,Renaud_2012}).

We have now demonstrated that we can observe a change in the probe
wave's travel-time induced by the shear-wave pump. In this section
we explore three aspects of how the elastic modulus of the rock changes
during the experiment. First, we look at the change in the modulus
$M=\lambda+2\mu$ induced by the pump. These results indicate that
we are not observing so-called slow-dynamics. As mentioned in the
introduction, rocks are known to exhibit a slow-dynamic response,
in which the rock is changed by a strong excitation (e.g. our pump)
and returns to its initial state slowly, over minutes to days (i.e.
over much longer time-scales than our measurements). Section~\ref{sub:Memory-free-measurements}
explores this phenomenon for our experimental setup. We then look
at changes in temperature, also known to affect nonlinear measurements.

\subsection{Nonlinear response of a Berea Sandstone}

\begin{figure}[h]
\includegraphics[width=0.5\textwidth]{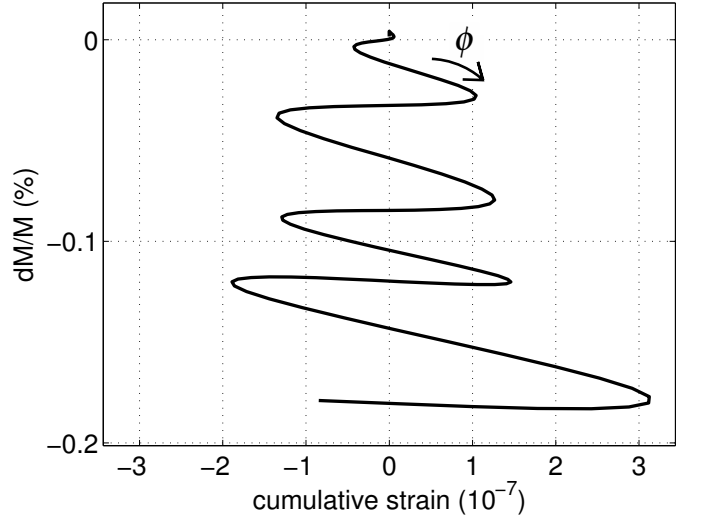} \protect\caption{\label{Fig 6}Strain dependency of the elastic modulus : the nonlinear
response of a Berea Sandstone is represented as a change in elastic
modulus $\nicefrac{\mathrm{d}M\left(\phi\right)}{M}$ in percent as
a function of the cumulative strain $\epsilon\left(\phi\right)=\int_{0}^{L}\epsilon\left(x,\phi+\frac{x}{c_{p}}\right)dx$
for each phase shift $\phi$ value from 0 ($\nicefrac{\mathrm{d}M\left(\phi\right)}{M}=0$)
to 140 ms ($\nicefrac{\mathrm{d}M\left(\phi\right)}{M}=-0.2\%$).
The quadratic non-linearity $\tilde{\beta}$ is responsible for the
oscillations and the global trend downward comes from the cubic component
$\tilde{\delta}$. }
\end{figure}

We first explore how the modulus changes with strain. Inverting Eq.~\ref{TOFM Moduli}
shows that the change in elastic modulus is directly related to the
relative time modulation as 
\begin{equation}
\frac{\mathrm{d}M\left(\phi\right)}{M}\approx-2\frac{TM\left(\phi\right)}{T_{o}}\,.\label{elastic modif}
\end{equation}
From this, we compute the change in modulus from the traveltime modulation.
To understand the relationship between this change in modulus and
the strain, induced by the pump, in the sample, we plot the left-hand
side of Eq.~\ref{elastic modif} as a function of the cumulative
pump strain 
\begin{equation}
\epsilon\left(\phi\right)=\int_{0}^{L}\epsilon\left(x,\phi+\frac{x}{c_{p}}\right)dx\,.\label{eq:cumstrain}
\end{equation}

It is now possible to represent the change of modulus $M$ as a function
of the strain. The maximum change is approximately 0.2\% of the elastic
modulus (c.f. \ref{tab:1}), which is similar to observations at large
scale during slow slip events \cite{Rivet_2011}, earthquakes \cite{Brenguier_2008}
or volcanism \cite{Brenguier_2008a}.

The curve shown in figure~\ref{Fig 6} shows a decrease in $M$ with
time ($\phi=0$ is the top of the plot), as well as an increase in
the cumulative pump strain. The quadratic term (Eq.~\ref{eq:Q})
is responsible for the small oscillations of the elastic modulus while
the cubic term (Eq.~\ref{eq:C}) explains this global weakening (decrease
of $M$) of the material. This weakening is a result of the accumulation
of strain over the time that the pump interacts with the sample, and
is not evidence of slow-dynamics. More precisely, in Eq.~\ref{eq:C}
we see that the cubic nonlinearity is governed by the integration
of $\epsilon^{2}$ along the path, over time. What we observe in Figure~\ref{Fig 6}
is the increase of this integral with time, i.e. as the pump continues
to oscillate within the sample it causes the integral of the square
of the strain (Eq.~\ref{eq:cumstrain}) to increase with time. Thus,
we are seeing an accumulation of the nonlinear effect as the pump
continues to propagate in the sample. In a perfect steady state regime,
each oscillation would describe the same curve. This regime is not
reached due to the finite size of the sample and the short recording
time.

\subsection{Absence of slow-dynamics\label{sub:Memory-free-measurements}}

\begin{figure}[h]
\includegraphics[width=0.5\textwidth]{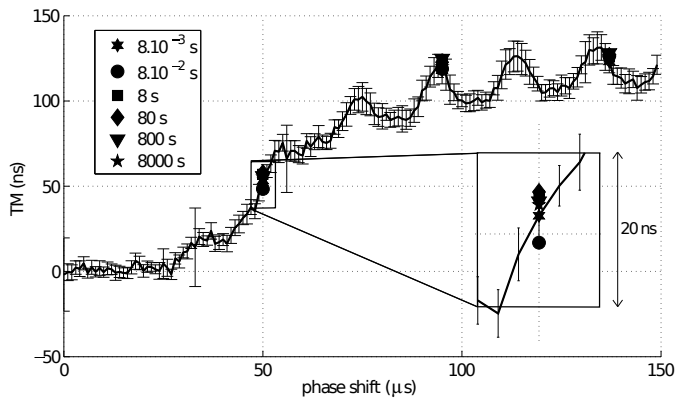} \protect\caption{{\tiny{}\label{fig:7}}The solid curve shows the time modulation $TM_{meas}$
measured with a delay $8\:\textrm{ms}$ between two pump activations
(i.e. the repetition frequency). This is the same curve as \ref{Fig 5}
b) with a different pump amplitude. An average over 10 acquisitions
was performed and the error bars reflect the variation between those
acquisitions. The time modulation is shown at 3 phase shift values
using delays of activation ranging from $8\:\textrm{ms}$ to 8000~s.
The inset gives a closer look at the different delays of acquisition
at one phase and shows that the variations for different delays is
within the noise of the experiment: there is no slow-dynamic effect
in the measurements.}
\end{figure}

As explained above, the change of elasticity shown in Figure~\ref{Fig 5}
is instantaneous. Nevertheless, as mentioned in the introduction,
it is known that the nonlinear response of rocks includes a memory
effect as reported by Holcomb for quasi-static measurement \cite{Holcomb_1981},
and by TenCate and others \cite{TenCate_1996,Johnson_2005} for dynamic
measurements. After a strong excitation they observed a weakening
that decreases with a very slow dynamic process, possibly lasting
up to several hours. This time scale can be explored in the present
experiment by changing the delay $d_{p}$ between two pump activations.
This means that we repeat the experiment at the same $\phi$ and vary
the wait time, $d_{p}$ between two experiments. We vary $d_{p}$
from 8~ms to 8000~s. The main curve of Fgure~\ref{fig:7} is another
acquisition of the time modulation already shown in Figure~\ref{Fig 5}
c), but with a higher pump amplitude. In order to limit the acquisition
time, the time modulation is measured at each $d_{p}$ for only 3
phase shifts values. Because of this limited measurement we cannot
estimate the nonlinear parameters. Nevertheless, it is clear in Figure~\ref{fig:7}
that measurements made with different values of the delay $d_{p}$,
all fall on the same curve, indicating that the delay does not affect
the time modulation $TM_{meas}$, and thus the nonlinear response.
It is possible that the maximum strain, on the order of a microstrain,
is too small to observe a slow dynamic process; this is of a similar
order-of-magnitude to that found in previous studies to cause slow
dynamics. For example, Pasqualini {\em et al} \cite{Pasqualini_2007}
report a threshold of around $5\times10^{-7}$ strain in sandstones
to produce a slow-dynamic effect. The main effect of the slow dynamic
process is a global weakening described by including a constant in
the elastic modulus to strain relationship given in equation \ref{aprox dm-1}.
Any measurement of $TM_{meas}$ is based on the acquisition of an
original probe and a perturbed probe when the pump is turned on (Eq.~\ref{Time modulation}),
which picks up a difference between two states and not the absolute
magnitude of the perturbation. As a consequence, if there is any slow
global weakening it is not measured here. Nevertheless, this observation
ensures that the measured parameters $\tilde{\beta}$ and $\tilde{\delta}$
are independent of the time properties of the acquisition sequence.

\subsection{Temperature effects}

\begin{figure*}[b]
\includegraphics[width=0.9\textwidth]{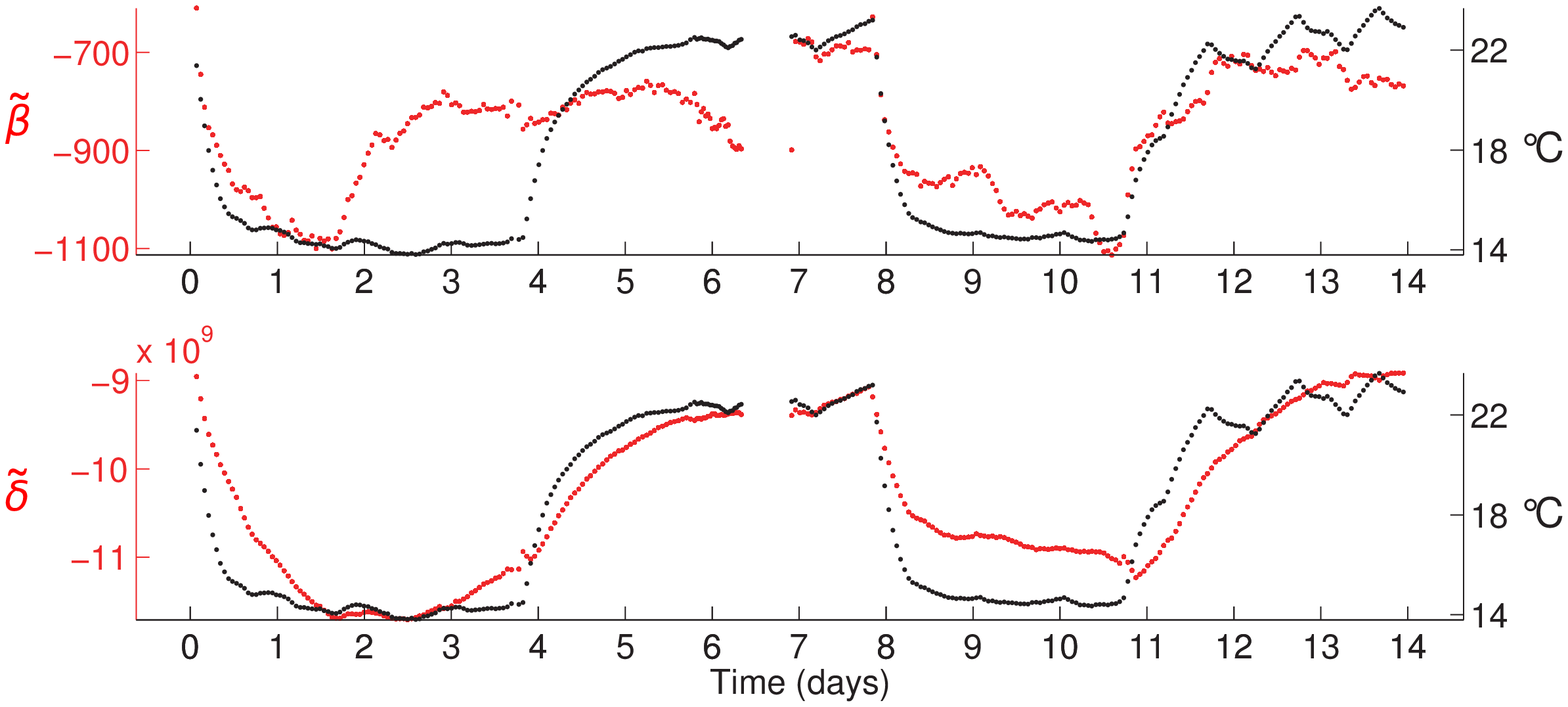}

\protect\caption{\label{fig:8}The 14-day evolution of $\tilde{\beta}$, $\tilde{\delta}$
(red dots) and the room temperature (in Celsius, black dots) demonstrates
the strong correlation between $\tilde{\delta}$ and the temperature,
and a fair correlation between $\tilde{\beta}$ and the temperature. }
\end{figure*}

The nonlinear response of materials are known to be sensitive to environmental
parameters. We measured $\tilde{\beta}$ and $\tilde{\delta}$ 300
times with a maximal strain of $2.5\times10^{-6}$ over a long period
of time (14 days) during which the room temperature switched two times
from the maximum to the minimum (from $22\,^{\circ}\mathrm{C}$ to
$15\,^{\circ}\mathrm{C}$) allowed by the room thermostat. The sample
was placed in a isothermal box in order to slow down the change in
temperature and damp the fluctuations; a thermometer was also placed
in the box to monitor the temperature. Figure~\ref{fig:8} shows
the evolution of $\tilde{\beta}$ and $\tilde{\delta}$ as a function
of the time and thus temperature. The maximum of the cross-correlation
between the cubic non-linearity $\tilde{\delta}$ and the temperature
is remarkable (0.91), and still important for $\tilde{\beta}$ (0.78).
The curious decorrelation in $\tilde{\beta}$ between day 2 and 6
probably involves other experimental parameters such as the humidity
and pressure that are also known to perturb the nonlinear response
\cite{Carpenter_2006,Ulrich_2001}.

Another unexplained feature of this experiment is the first measurement
of $\tilde{\beta}$ after half a day with no pump on day 7 which is
50\% smaller than subsequent points. This bias was noted for all the
experiments and for this reason, the first measurement is excluded
when an average is performed. This phenomenon is probably related
to a slow-dynamic effect with a very long recovery time as it is not
observed after 2.2~hours (8000~s) in the previous experiment (Figure~\ref{fig:7}).
Both the effects of temperature decorrelation and first acquisition
bias are only present on $\tilde{\beta}$ demonstrating that $\tilde{\beta}$
and $\tilde{\delta}$ are independent and likely have different physical
origins. The following section discusses the physical meaning of these
nonlinear parameters.

In making these measurements, we were not able to directly monitor
the strain amplitude as the laser vibrometer was not available. Estimates
of strain are based on the application of the same voltage to the
pump transducer as used in previous experiments. Our goal here, is
to characterize the effect that varying temperature has on our results.
It remains possible that the origin of this effect is not a change
in the nonlinearity of the rock itself, but instead a change in the
apparatus resulting in a change in the induced pump strain. It is
clear, however, that the effect of temperature cannot be ignored;
to mitigate this effect in other experiments, we use a combination
of shielding to reduce temperature fluctuations and speed to make
the measurements as quickly as possible to avoid the effects of such
fluctuations on the result. Note also that the large fluctuations
shown in Figure~\ref{fig:8} are a result of a $\sim10\,^{\circ}\mathrm{C}$
fluctuation which is significantly more than is usually observed in
our laboratory.

\subsection{Strain amplitude dependency\label{sub:Strain-amplitude-dependency}}

\begin{figure}[H]
\includegraphics[width=0.5\textwidth]{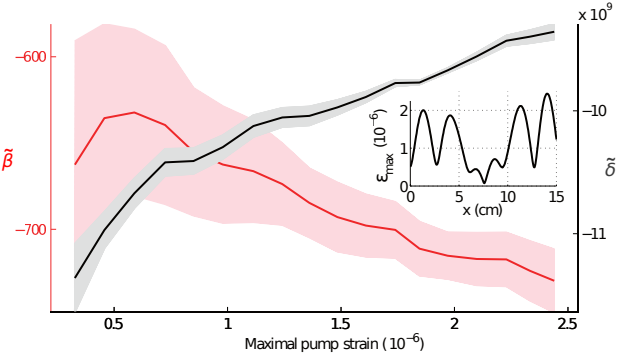}

\protect\caption{\label{fig:9}$\tilde{\beta}$ and $\tilde{\delta}$ (respectively
red and black) as a function of the maximum induced strain in the
sample by different amplitudes of the shear wave pump. Above 0.6 microstrain,
$\tilde{\beta}$ decreases with strain whereas $\tilde{\delta}$ increases
implying different mechanisms. The abscissa is the maximum of the
strain over the whole propagation time of the pump measured along
the probe path ($0<x<15$~cm, $0<y<3$~cm, $1.5<z<4.5$~cm). The
variation of this maximal strain along the $x$-axis shown in the
inset illustrates the inhomogeneous spatial distribution of strain
in the sample, obtained from the finite difference simulation.}
\end{figure}

In the description of the nonlinear parameters given in section~\ref{sec:TMtoelastic},
we discuss a nonlinear Hooke's law. This can, of course, be translated
to a wave equation in which the wave speed becomes strain dependent
(cf Eq.~\ref{eq: fourth_order_elasticity}). As a result, the wave
propagation becomes strain (or equivalently source amplitude) dependent.
For rocks it is reported that the strain wave-speed relationship is
itself amplitude dependent meaning that the nonlinear parameters depend
on the strain \cite{Renaud_2013}. In other words, the nonlinear response
of rocks depends on the maximum strain. The nonlinear elastic model
is thua valid only at a fixed maximum strain of the pump, and the
nonlinear coefficients $\tilde{\beta}$ and $\tilde{\delta}$ are
functions of this amplitude.

Our experimental set-up enables the characterization of this feature
of the nonlinearities. Estimates of the nonlinear parameters were
repeated for 18 pump shear wave amplitudes. The induced strain along
the probe path, estimated by the method described in section~\ref{sub:Quantifying-the-pump-strain},
attains a maximum ranging from 0.3 to 2.2~microstrain. Figure~\ref{fig:9}
shows $\tilde{\beta}$ and $\tilde{\delta}$ as a function of the
strain and their standard deviation among 300 sets of 18 pump amplitudes.
Each set represents one hour of acquisition. The averaging over 300
acquisitions is performed after an adjustment of the median value
of each acquisition set in order to remove the environmental effects
such as the temperature effects discussed above. The standard deviation
is clearly related to the signal to noise ratio as it decreases with
increasing pump strain and is much bigger for $\tilde{\beta}$, whose
estimation is based on a signal approximately 7 times smaller than
that of $\tilde{\delta}$. Figure~\ref{fig:9} demonstrates that
above a microstrain $\tilde{\delta}$ increases linearly with the
strain and $\tilde{\beta}$ decreases linearly with it. The changes
are noticeable but remain small for nonlinearities (less than 20\%).
The quadratic nonlinearity decreases with the absolute strain while
the cubic nonlinearity, which is primarily responsible for the rock
softening, increases. This is in agreement with the observations of
Renaud {\em et al.} \cite{Renaud_2013,Renaud_2013a}. The different
strain dependencies for the 2 nonlinear parameters suggests that the
underlying mechanisms from which the quadratic and cubic nonlinearity
originate are different.

The inset in figure~\ref{fig:9} shows the spatial distribution of
the strain along the probe wave path, modeled with finite difference
and scaled to the experiment as described above, and clearly shows
that the strain in the medium is not homogeneous. The antisymmetry
axis at $x=7.5$~cm is in agreement with the wave response (in strain)
to a point force: compression in one half space and tension in the
other one. The free boundaries conditions at $x=0$ and $x=15$~cm
also have a clear effect on the spatial distribution of strain.

Because of this antisymmetry the maximal strain is only an indicator
of the strain amplitude it has to be interpreted carefully when comparing
to methods where the strain is nearly uniform. Furthermore, since
the nonlinear parameters are amplitude dependent, the spatial distribution
of the strain may also affect the measurement, even with the spatio-temporal
integration described in Eq.~\ref{TOFM Moduli-1}.

\subsection{Pump orientation}

\begin{figure}[H]
\includegraphics[width=0.5\textwidth]{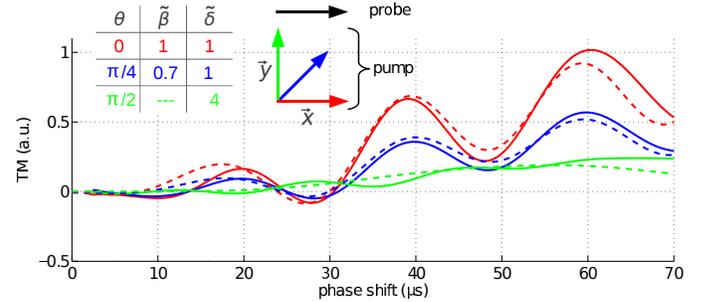}\protect\caption{\label{fig:10} The measured (solid lines) and modeled (dashed lines)
time modulations are shown for different values of the angle $\theta$
between the $x$-axis and the orientation of the S-wave transducer
$\protect\overrightarrow{p}$. The main pump direction of the strain
is indicated by the colored arrows and the probe direction is in black.
The red line correspond to the standard case where probe and pump
are along the $x$-axis in red (particle motions are aligned). The
normalized nonlinear parameters $\tilde{\beta}$ and $\tilde{\delta}$
in the inset show the anisotropy of the nonlinear response.}
\end{figure}

The shear wave pump creates an anisotropy in the medium, as any uni-axial
static load would do \cite[p. 64]{Murnaghan_1951}. In other words,
the mechanical response of an isotropic medium becomes dependent on
the direction of the pump; this effect vanishes when the pump is turned
off. In this section we study this effect by changing the direction
of the strain pump shear wave particle motion relative to the probe
direction. Previously, the propagation of the probe wave and the pump
strain occurred in the $x-$direction in the $xz$-plane (Figure~\ref{fig:1}),
i.e. the particle motions of the pump and probe are aligned. It is
convenient to change the pump strain direction with a rotation of
the pump transducer about the $z-$axis (directed downward in Figure~\ref{fig:1}).
This rotation is a technical challenge of maintaining constant coupling
between the S-wave transducer and the sample. We solve this by applying
a homogeneous and perfectly oriented force (along the $z-$axis) to
the transducer. This force should create a constant coupling, insensitive
to a rotation around the $z-$axis. We find the best solution to be
a cylindrical load above the transducer with a significant layer of
S-wave couplant to provide adequate lubrication during rotation and
to minimize the variation in coupling due to drying of the couplant.
The rotation of the transducer was carefully performed by small steps
in angle to minimize any other perturbation from the change in weight
distribution. The stability of the method was checked visually with
a transparent sample and quantitatively as described below.

The $x-$component of the displacement was measured with another shear
wave transducer placed at the position of the laser beam in Fig. \ref{fig:1}:
3 cm under the pump transducer on the $\left(\overrightarrow{z},\overrightarrow{y}\right)$
surface. The measurements of the $x-$component of the displacement
for the pump transducer oriented along $\overrightarrow{p}=\mathrm{cos}\theta\:\overrightarrow{x}+\mathrm{sin}\theta\:\overrightarrow{y}$
were found to be close to the projection of $\overrightarrow{p}$
along $\overrightarrow{x}$, within a 10\% error. This indicates that
the coupling remains relatively constant during the rotation of the
pump transducer.

The nonlinear elastic model in Eq.~\ref{aprox dm-1} do not includes
the $\epsilon_{yz}$ because we noted that this term was negligible
(see Eq.~\ref{relative strain amplitude}). When $\theta=\frac{\pi}{2}$,
the main component of the displacement is along $y-$axis and $\epsilon_{yz}$
becomes the main strain component. Including this term in the elastic
model modifies equation \ref{aprox dm-1} as follows:

\begin{align}
\frac{\mathrm{d}M}{M}&\approx\tilde{\beta}\left(\epsilon_{xx}+\frac{2}{5}\epsilon_{yy}+\frac{2}{5}\epsilon_{zz}\right) \nonumber\\
+&\tilde{\delta}\left(\epsilon_{xx}^{2}+\frac{5}{12}\epsilon_{yy}^{2}+\frac{5}{12}\epsilon_{zz}^{2}+\frac{3}{32}\epsilon_{xz}^{2}+\frac{1}{16}\epsilon_{yz}^{2}\right)\,.\label{aprox dm-1-1}
\end{align}
Then, the same procedure described in section~\ref{sub:Quantifying-the-pump-strain}
is performed to estimate $\tilde{\beta}$ and $\tilde{\delta}$ as
a function of the angle $\theta$. The finite difference simulation
was performed for $\theta=[0,\frac{\pi}{4},\frac{\pi}{2}]$ in order
to estimate the strain components within the sample and compute the
quantities defined in Eq.~\ref{eq:Q} and \ref{eq:C}, with the only
change of $\epsilon^{2}=\epsilon_{xx}^{2}+\frac{5}{12}\epsilon_{yy}^{2}+\frac{5}{12}\epsilon_{zz}^{2}+\frac{3}{32}\epsilon_{xz}^{2}+\frac{1}{16}\epsilon_{yz}^{2}$.
The nonlinear parameters $\tilde{\beta}$ and $\tilde{\delta}$ are
estimated from the measured time modulation using Eq.~\ref{eq:beta_aprox}
and \ref{eq:delta_aprox}. Because no laser measurements were available
in this experiment, only the relative value of strain is estimated
and the nonlinear parameters are normalized by their value at $\theta=0$.

The measured and simulated time modulations are shown in figure~\ref{fig:10}
and establish the dependency of the nonlinear parameters on the angle
$\theta$. No value of $\tilde{\beta}$ is at $\theta=\frac{\pi}{2}$
because the measured fast component of $TM_{meas}$ is very close
to the noise level and does not have any phase correlation with the
modeled signal. Nevertheless, the value of $\tilde{\beta}(\frac{\pi}{4})=0.7$
indicates that the quadratic nonlinearity decreases when the direction
of the particle motion of the pump is orthogonal to that of the probe.
On the contrary $\tilde{\delta}(\frac{\pi}{2})=4$ shows an increase
of the cubic nonlinearity in this case. This apparent anisotropy has
to be considered carefully because, $\tilde{\beta}$ and $\tilde{\delta}$
may vary with the strain distribution (see section~\ref{FOET}).
This distribution clearly changes when the pump transducer is rotated
and this may bias the anisotropy measurement.

\section{Conclusion}

Previous experiments of nonlinear elastic effects depended on standing
waves and finite-sized samples under compressionnal stress. In this
work, we demonstrate the feasibility of using two propagating waves
for estimating nonlinear properties of a rock. In our experiments,
a microstrain pump wave modulates a probe wave; the resulting arrival
time modulation was determined to be a cubic function of the complex
strain field. The measured time modulation is on the order of tens
of nanoseconds, measured in a Berea Sandstone sample with a 50~kHz
S-wave pump and a 0.5~MHz P-wave probe. We fit the time modulation
data with a two-parameter model: a quadratic and a cubic nonlinearity
term related theoretically to averaged elastic moduli of third and
fourth order respectively. Temperature, strain amplitude, and the
polarization of the pump wave relative to the probe wave direction
can affect the measured time delays; longer term slow-dynamic effects
do not appear to be significant. Future work will be directed towards
investigating larger samples and different types of rocks.

\section{Acknowledgments}

The authors would like to thank Weatherford, as well as the Earth
Resources Laboratory, for funding this research. We are also grateful
to Xinding Fang for helping us to setup the modeling code. We thank
Jim TenCate and an anonymous reviewer for helpful comments that greatly
improved the clarity of the paper.

\bibliographystyle{unsrt}

\begin{thebibliography}{10}

\bibitem{Belyayeva_1995}
IY~Belyayeva, VY~Zaitsev, and AM~Sutin.
\newblock Tomography of nonlinear rock parameters of seismology and seismic
  prospecting.
\newblock {\em {Fizika Zemli}}, ({12}):{44--51}, {DEC} {1995}.

\bibitem{Guyer_1999}
RA~Guyer and PA~Johnson.
\newblock {Nonlinear mesoscopic elasticity: Evidence for a new class of
  materials}.
\newblock {\em Physics Today}, {52}({4}):{30--36}, {APR} {1999}.

\bibitem{Guyer_2009}
Robert~A Guyer and Paul~A Johnson.
\newblock Nonlinear mesoscopic elasticity.
\newblock 2009.

\bibitem{Regnier_2013}
J.~Regnier, L.F. Cadet, H.and~Bonilla, E.~Bertrand, and J.F. Semblat.
\newblock {Assessing Nonlinear Behavior of Soils in Seismic Site Response:
  Statistical Analysis on KiK-net Strong-Motion Data}.
\newblock {\em Bulletin of the Seismological Society of America},
  {103}({3}):{1750--1770}, {JUN} {2013}.

\bibitem{Brenguier_2008}
F.~Brenguier, M.~Campillo, C.~Hadziioannou, N.~M. Shapiro, R.~M. Nadeau, and
  E.~Larose.
\newblock {Postseismic relaxation along the San Andreas fault at Parkfield from
  continuous seismological observations}.
\newblock {\em Science}, {321}({5895}):{1478--1481}, {SEP 12} {2008}.

\bibitem{Kurtulus_2008}
A~Kurtulus and KH~Stokoe.
\newblock In situ measurement of nonlinear shear modulus of silty soil.
\newblock {\em Journal of geotechnical and geoenvironmental engineering},
  134(10):1531--1540, 2008.

\bibitem{Johnson_2009}
P.A. Johnson, P.~Bodin, J.~Gomberg, F.~Pearce, Z.~Lawrence, and F.Y. Menq.
\newblock {Inducing in situ, nonlinear soil response applying an active
  source}.
\newblock {\em Journal of Geophysical Research: Solid Earth}, {114}, {MAY 21}
  {2009}.

\bibitem{Lawrence_2009}
Z.~Lawrence, P.~Bodin, and Ch.A. Langston.
\newblock {In Situ Measurements of Nonlinear and Nonequilibrium Dynamics in
  Shallow, Unconsolidated Sediments}.
\newblock {\em Bulletin of the Seismological Society of America},
  {99}({3}):{1650--1670}, {JUN 1} {2009}.

\bibitem{Cox_2009}
B.R. Cox, K.H. Stokoe, II, and E.M. Rathje.
\newblock {An In Situ Test Method for Evaluating the Coupled Pore Pressure
  Generation and Nonlinear Shear Modulus Behavior of Liquefiable Soils}.
\newblock {\em {Geotechnical Testing Journal}}, {32}({1}):{11--21}, {JAN}
  {2009}.

\bibitem{Holcomb_1981}
David~J Holcomb.
\newblock Memory, relaxation, and microfracturing in dilatant rock.
\newblock {\em Journal of Geophysical Research: Solid Earth},
  86(B7):6235--6248, 1981.

\bibitem{Thomsen_1995}
L.~Thomsen.
\newblock Elastic anisotropy due to aligned cracks in porous rock.
\newblock {\em Geophysical Prospecting}, {43}({6}):{805--829}, {AUG} {1995}.

\bibitem{SAYERS_1995}
CM~SAYERS and M~KACHANOV.
\newblock Microcrack-induced elastic wave anisotropy of brittle rocks.
\newblock {\em Journal of Geophysical Research: Solid Earth},
  {100}({B3}):{4149--4156}, {MAR 10} {1995}.

\bibitem{Gueguen_2003}
Y~Gueguen and A~Schubnel.
\newblock {Elastic wave velocities and permeability of cracked rocks}.
\newblock {\em Tectonophysics}, {370}({1-4}):{163--176}, {JUL 31} {2003}.

\bibitem{Fortin_2007}
J.~Fortin, Y.~Gueguen, and A.~Schubnel.
\newblock {Effects of pore collapse and grain crushing on ultrasonic velocities
  and V-p/V-s}.
\newblock {\em Journal of Geophysical Research: Solid Earth}, {112}({B8}), {AUG
  21} {2007}.

\bibitem{JOHNSON_1991}
P.~A Johnson, A.~Migliori, and T.J. Shankland.
\newblock Continuous wave phase detection for probing nonlinear elastic wave
  interactions in rocks.
\newblock {\em Journal of the Acoustical Society of America},
  {89}({2}):{598--603}, {1991}.

\bibitem{TenCate_1996}
JA~TenCate and TJ~Shankland.
\newblock {Slow dynamics in the nonlinear elastic response of Berea sandstone}.
\newblock {\em Geophysical Research Letters}, {23}({21}):{3019--3022}, {OCT 15}
  {1996}.

\bibitem{Gusev_1998}
VE~Gusev, W~Lauriks, and J~Thoen.
\newblock {Dispersion of nonlinearity, nonlinear dispersion, and absorption of
  sound in micro-inhomogeneous materials}.
\newblock {\em Journal of the Acoustical Society of America},
  {103}({6}):{3216--3226}, {JUN} {1998}.

\bibitem{DAngelo_2004}
RM. D'Angelo, KW. Winkler, TJ. Plona, BJ. Landsberger, and DL. Johnson.
\newblock {Test of hyperelasticity in highly nonlinear solids: Sedimentary
  rocks}.
\newblock {\em Physical Review Letters}, {93}({21}), {2004}.

\bibitem{Darling_2004}
T.W. Darling, J.A. TenCate, D.W. Brown, B.~Clausen, and S.C. Vogel.
\newblock {Neutron diffraction study of the contribution of grain contacts to
  nonlinear stress-strain behavior}.
\newblock {\em Geophysical Research Letters}, {31}({16}), {AUG 26} {2004}.

\bibitem{Renaud_2008}
G.~Renaud, S.~Calle, J.P. Remenieras, and M.~Defontaine.
\newblock {Exploration of trabecular bone nonlinear elasticity using
  time-of-flight modulation}.
\newblock {\em IEEE Transactions on Ultrasonics Ferroelectrics and Frequency
  Control}, {55}({7}):{1497--1507}, {2008}.
\newblock {IEEE Ultrasonics Symposium, New York, NY, OCT 28-31, 2007}.

\bibitem{Renaud_2009}
G.~Renaud, S.~Calle, and M.~Defontaine.
\newblock {Remote dynamic acoustoelastic testing: Elastic and dissipative
  acoustic nonlinearities measured under hydrostatic tension and compression}.
\newblock {\em Applied Physics Letters}, {94}({1}), {JAN 5} {2009}.

\bibitem{Muller_2006}
M.~Muller, A.~Sutin, R.~Guyer, M.~Talmant, P.~Laugier, and P.A. Johnson.
\newblock Nonlinear resonant ultrasound spectroscopy (nrus) applied to damage
  assessment in bone.
\newblock {\em Journal of the Acoustical Society of America}, 118:3946, 2005.

\bibitem{Renaud_2012}
G.~Renaud, P.~Y. Le~Bas, and P.~A. Johnson.
\newblock {Revealing highly complex elastic nonlinear (anelastic) behavior of
  Earth materials applying a new probe: Dynamic acoustoelastic testing}.
\newblock {\em Journal of Geophysical Research: Solid Earth}, {117}, {JUN 6}
  {2012}.

\bibitem{Renaud_2013}
G.~Renaud, J.~Riviere, S.~Haupert, and P.~Laugier.
\newblock {Anisotropy of dynamic acoustoelasticity in limestone, influence of
  conditioning, and comparison with nonlinear resonance spectroscopy}.
\newblock {\em Journal of the Acoustical Society of America},
  {133}({6}):{3706--3718}, {JUN} {2013}.

\bibitem{Renaud_2013a}
G.~Renaud, J.~Riviere, P.~Y. Le~Bas, and P.~A. Johnson.
\newblock {Hysteretic nonlinear elasticity of Berea sandstone at
  low-vibrational strain revealed by dynamic acousto-elastic testing}.
\newblock {\em Geophysical Research Letters}, {40}({4}):{715--719}, {FEB 28}
  {2013}.

\bibitem{Riviere_2013}
J~Riviere, G~Renaud, RA~Guyer, and PA~Johnson.
\newblock Pump and probe waves in dynamic acousto-elasticity: Comprehensive
  description and comparison with nonlinear elastic theories.
\newblock {\em Journal of Applied Physics}, 114(5):054905--054905, 2013.

\bibitem{Renaud_2013b}
G~Renaud, J~Rivi{\`e}re, C~Larmat, JT~Rutledge, RC~Lee, RA~Guyer, K~Stokoe, and
  PA~Johnson.
\newblock In situ characterization of shallow elastic nonlinear parameters with
  dynamic acoustoelastic testing.
\newblock {\em Journal of Geophysical Research: Solid Earth}, 2014.

\bibitem{Geza_2001}
N.~Geza, G.~Egorov, Y.~Mkrtumyan, and V.~Yushin.
\newblock Instantaneous variations in velocity and attenuation of seismic waves
  in a friable medium in situ under pulsatory dynamic loading: An experimental
  study,.
\newblock {\em Russ. Geol. Geophys.}, 42(7):1079--1087, 2001.

\bibitem{Toupin_1961}
R.~A. Toupin and B.~Bernstein.
\newblock Sound waves in deformed perfectly elastic materials. acoustoelastic
  effect.
\newblock {\em The Journal of the Acoustical Society of America},
  33(2):216--225, 1961.

\bibitem{Johnson_1989}
Paul~A Johnson and Thomas~J Shankland.
\newblock Nonlinear generation of elastic waves in granite and sandstone:
  continuous wave and travel time observations.
\newblock {\em Journal of Geophysical Research: Solid Earth (1978--2012)},
  94(B12):17729--17733, 1989.

\bibitem{Destrade_2010}
Michel Destrade, Michael~D. Gilchrist, and Raymond~W. Ogden.
\newblock Third- and fourth-order elasticities of biological soft tissues.
\newblock {\em The Journal of the Acoustical Society of America},
  127(4):2103--2106, 2010.

\bibitem{Destrade_2010a}
Michel Destrade, Michael~D. Gilchrist, and Giuseppe Saccomandi.
\newblock Third- and fourth-order constants of incompressible soft solids and
  the acousto-elastic effect.
\newblock {\em The Journal of the Acoustical Society of America},
  127(5):2759--2763, 2010.

\bibitem{Destrade_2010b}
Michel Destrade and Raymond~W. Ogden.
\newblock On the third- and fourth-order incompressible isotropic elasticity.
\newblock {\em The Journal of the Acoustical Society of America},
  128(6):3334--3343, 2010.

\bibitem{Johnson_2005}
P~Johnson and A~Sutin.
\newblock {Slow dynamics and anomalous nonlinear fast dynamics in diverse
  solids}.
\newblock {\em Journal of the Acoustical Society of America},
  {117}({1}):{124--130}, {2005}.

\bibitem{Catheline_1999}
S.~Catheline, F.~Wu, and M.~Fink.
\newblock A solution to diffraction biases in sonoelasticity: The acoustic
  impulse technique.
\newblock {\em Journal of the Acoustical Society of America}, 105
  (5):2941--2950, 1999.

\bibitem{Winkler_1996}
Kenneth~W Winkler and Xingzhou Liu.
\newblock Measurements of third-order elastic constants in rocks.
\newblock {\em The Journal of the Acoustical Society of America},
  100(3):1392--1398, 1996.

\bibitem{Virieux_1986}
Jean Virieux.
\newblock P-sv wave propagation in heterogeneous media: Velocity-stress
  finite-difference method.
\newblock {\em Geophysics}, 51(4):889--901, 1986.

\bibitem{Graves_1996}
Robert~W Graves.
\newblock Simulating seismic wave propagation in 3d elastic media using
  staggered-grid finite differences.
\newblock {\em Bulletin of the Seismological Society of America},
  86(4):1091--1106, 1996.

\bibitem{Landau_1986}
LD~Landau and EM~Lifshitz.
\newblock {\em Course of Theoretical Physics Vol 7: Theory of Elasticity}.
\newblock Pergamon Press, Oxford, UK, 3rd edition, 1986.

\bibitem{Zabolotskaya_1986}
EA~Zabolotskaya.
\newblock Sound beams in a nonlinear isotropic solid.
\newblock {\em SOVIET PHYSICS ACOUSTICS-USSR}, 32(4):296--299, 1986.

\bibitem{Jacob_2007}
Xavier Jacob, Stefan Catheline, Jean-Luc Gennisson, Christophe Barriere, Daniel
  Royer, and Mathias Fink.
\newblock {Nonlinear shear wave interaction in soft solids}.
\newblock {\em Journal of the Acoustical Society of America},
  {122}({4}):{1917--1926}, {2007}.

\bibitem{Abiza_2012}
Zaki Abiza, Michel Destrade, and Ray~W Ogden.
\newblock Large acoustoelastic effect.
\newblock {\em Wave Motion}, 49(2):364--374, 2012.

\bibitem{Hamilton_1998}
M.F. Hamilton, D.T. Blackstock, et~al.
\newblock {\em Nonlinear acoustics}.
\newblock Academic press., 1998.

\bibitem{Renaud_2011}
G.~Renaud, M.~Talmant, S.~Calle, M.~Defontaine, and P.~Laugier.
\newblock {Nonlinear elastodynamics in micro-inhomogeneous solids observed by
  head-wave based dynamic acoustoelastic testing}.
\newblock {\em Journal of the Acoustical Society of America},
  {130}({6}):{3583--3589}, {DEC} {2011}.

\bibitem{Rivet_2011}
D.~Rivet, M.~Campillo, N.M. Shapiro, V.~Cruz-Atienza, M.~Radiguet, N.~Cotte,
  and V.~Kostoglodov.
\newblock {Seismic evidence of nonlinear crustal deformation during a large
  slow slip event in Mexico}.
\newblock {\em Geophysical Research Letters}, {38}, {APR 28} {2011}.

\bibitem{Brenguier_2008a}
F.~Brenguier, N.M. Shapiro, M.~Campillo, V.~Ferrazzini, Z.~Duputel, O.~Coutant,
  and A.~Nercessian.
\newblock {Towards forecasting volcanic eruptions using seismic noise}.
\newblock {\em Nature Geoscience}, 1(2):126--130, 2008.

\bibitem{Pasqualini_2007}
D.~Pasqualini, K.~Heitmann, J.A. TenCate, S.~Habib, D.d Higdon, and P.A.
  Johnson.
\newblock {Nonequilibrium and nonlinear dynamics in Berea and Fontainebleau
  sandstones: Low-strain regime}.
\newblock {\em Journal of Geophysical Research: Solid Earth}, {112}({B1}), {JAN
  23} {2007}.

\bibitem{Carpenter_2006}
M.A. Carpenter, P.~Sondergeld, B.~Li, R.C. Liebermann, J.W. Walsh, J.~Schreuer,
  and T.W. Darling.
\newblock {Structural evolution, strain and elasticity of perovskites at high
  pressures and temperatures}.
\newblock {\em Journal of Mineralogical and Petrological Sciences},
  {101}({3}):{95--109}, {JUN} {2006}.

\bibitem{Ulrich_2001}
TJ~Ulrich and TW~Darling.
\newblock {Observation of anomalous elastic behavior in rock at low
  temperatures}.
\newblock {\em Geophysical Research Letters}, {28}({11}):{2293--2296}, {JUN 1}
  {2001}.

\bibitem{Murnaghan_1951}
F.D. Murnaghan.
\newblock {\em Finite Deformation of an elastic solid}.
\newblock New york John wiley \& sons, inc., 1951.

\end{thebibliography}

\end{document}